\documentclass{bytedance}

\usepackage[utf8]{inputenc}
\usepackage{amsmath}
\usepackage{amssymb}
\usepackage{amsfonts}
\usepackage{adjustbox}
\usepackage{tabularx}
\usepackage{nicefrac}
\usepackage{newfloat}
\usepackage{listings}
\usepackage{pifont}

\DeclareRobustCommand{\swanicon}{\raisebox{-0.15em}{\includegraphics[height=1.1em]{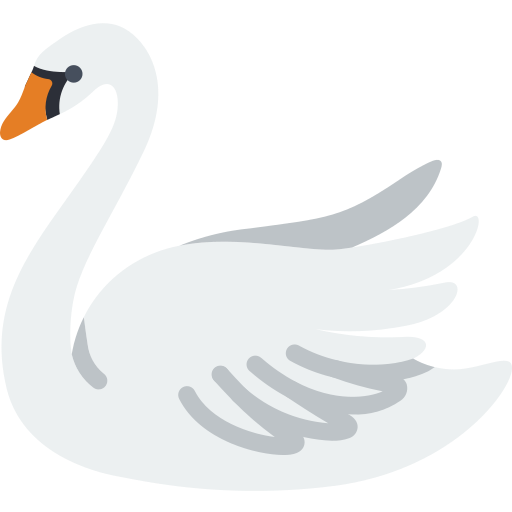}}}

\AtBeginEnvironment{tabular}{\small}
\AtBeginEnvironment{tabularx}{\small}

\definecolor{darkorange}{rgb}{1.0, 0.55, 0.0}
\definecolor{mygreen}{RGB}{0,180,0}
\definecolor{myred}{RGB}{180,0,0}

\title{\swanicon\ SwanTale: Unified Multi-Speaker Speech and Audio Generation for Instruct and Zero-Shot Tasks}

\author[1,*]{Yu Zhang}
\author[1,*]{Ruiqi Li}
\author[1,2,*]{Changhao Pan}
\author[1,2]{Ke Lei}
\author[1,\dagger]{Xiang Yin}
\author[1]{Cheng Yang}

\affiliation[1]{ByteDance}
\affiliation[2]{Zhejiang University}

\contribution[*]{Equal contribution}
\contribution[\dagger]{Corresponding author}
\correspondence{\href{mailto:zhangyu.34@bytedance.com,liruiqi.23@bytedance.com,yinxiang.stephen@bytedance.com}{\texttt{\{zhangyu.34,liruiqi.23,yinxiang.stephen\}@bytedance.com}}}
\project{\url{https://swanaigc.github.io/\#swantale}}

\abstract{
Speech and audio generation is often needed in animation dubbing, audio drama, movies, advertising, games, podcasts, and short-video production. In these scenarios, creators may need to design voices without reference recordings, control speaker styles with natural language, support acoustic scenes with environments and audio effects, and later reuse the designed voices. Therefore, it is important to support multi-speaker speech and audio generation for both instruct and zero-shot tasks. The instruct task requires a caption of the environment, speaker styles, and fine-grained content, while the zero-shot task uses reference audio together with the same fine-grained content. We address these tasks from both the data and model sides. First, we propose \textbf{SwanData-Caption}, which cleans raw speech and audio data, adds targeted synthetic coverage, and annotates diverse and accurate multi-level captions. Then, we propose \textbf{SwanTale}, a multi-speaker expressive speech and audio generation model that supports both zero-shot and instruct tasks. We introduce SwanVAE to support high-quality multi-audio-modality generation. Then, we adopt reward-conditioned quality control and Engram conditioning, along with Unified MoE for multi-task and multi-audio-modality modeling. In addition, we use curriculum learning and GRPO post-training to let the model progressively learn and strengthen its capabilities. Experimental results show that SwanTale leads on multiple key zero-shot and instruct metrics, achieves the best expressiveness scores in both tasks, and supports complex instruct generation involving multi-speaker speech and audio. 
}

\begin{document}

\maketitle

\section{Introduction}
\label{sec:intro}

Recent text-to-speech (TTS) systems have made substantial progress in zero-shot speech synthesis. Given only a short reference audio clip, these systems can synthesize vivid new content according to the reference voice \citep{chen2024f5,du2024cosyvoice,shen2023naturalspeech,jiang2024mega,huang2022generspeech}.
This capability has made speaker voice cloning and multi-speaker speech generation practical when the desired voice already exists in a recording \citep{zhang2024covomix,ju2025mooncast,zhu2025zipvoice,li2026swanvoice,xie2025fireredtts,zhang2025conan}. Media creation, however, often begins from the opposite case. In animation dubbing, audio drama, movies, advertising, games, podcasts, and short-video production, the target speaker may not have a reference recording at all. Creators need to design a voice from scratch, specify how the character should perform a line, and place that line inside an acoustic scene that changes how the performance is perceived; later, the designed voice may be reused through zero-shot synthesis. A modern TTS system for such workflows should therefore support both the zero-shot task with reference audio and the instruct task controlled by natural language captions.

In the classical zero-shot speech synthesis task, the input is speech content paired with reference audio. The reference audio provides speaker identity. In addition to speech content and, for dialogue tasks, speaker-turn labels, recent systems increasingly use local style descriptions to control emotion and speaking rate \citep{chen2026flexivoice,du2025cosyvoice}. By contrast, in the instruct speech synthesis task, the input is only a caption, which may describe the environment, speaker styles, and fine-grained content. Here, \emph{environment} includes scene type, sound-field or recording-space cues, and persistent background audio effects \citep{huang2023make,zhu2025asaudio,pan2025spatialeval}. \emph{Speaker styles} include not only gender and age, but also persona, role, timbre, habitual style, and other stable voice traits. \emph{Fine-grained content} covers the information in additional local style descriptions and can also describe local audio effects \citep{zhou2025indextts2,zhang2024tcsinger,guo2025techsinger}. Existing instruct TTS systems can already control dimensions such as speaker age, gender, emotion, and pace \citep{lacombe2024parlertts,huang2025instructttseval,chen2026flexivoice,ren2026ovinstructtts,hu2026voicesculptor}. Yet most of them still generate only speech. For scenarios that require environment and local audio effects, if these components are produced by a downstream audio pipeline, their timing, loudness, reverberation, and acoustic texture can drift away from the speech even when each component is plausible in isolation. Existing designs also often reduce speaker control to decomposed attribute labels, while creators may want a persona description in natural language to map directly to a voice. Finally, existing instruct TTS systems are usually not designed to preserve zero-shot capability in the same model. Therefore, supporting both long-form multi-speaker zero-shot and highly expressive instruct speech generation with environment, speaker personas, fine-grained content, and audio effects remains an open problem.

Overall, this task faces three main challenges.
\textbf{(1) Data scarcity.} Zero-shot TTS can rely on public speech corpora \citep{zen2019libritts,chen2021gigaspeech,kang2024libriheavy,shi2020aishell,panayotov2015librispeech}, but diverse and high-quality caption data for the instruct task requires richer audio coverage and more detailed caption annotation. Collecting expressive speech data, as well as clean audio data, is costly \citep{yang2024realman,zhang2025isdrama,guo2025mrsaudio}, and annotating multi-level natural-language captions is also expensive \citep{xu2024secap,zhang2024gtsinger,guo2025stars}.
\textbf{(2) Task compatibility.} Instruct samples describe speaker styles through natural-language captions, whereas zero-shot samples obtain speaker styles from reference audio. At the same time, both tasks need to share fine-grained content captions. Joint training must preserve the shared speech modeling while preventing the two conditioning paths from weakening each other \citep{qwen2026qwen3tts,zhang2025tcsinger}.
\textbf{(3) Multi-audio-modality complexity.} Our task aims to generate expressive speech, general audio, occasional singing voice, and music within a single waveform, while preserving intelligibility and speaker identity \citep{yang2023uniaudio,liu2025unimoeaudio,lei2026swansphere}. These components have different temporal structures: speech needs lexical alignment and stable identity, environmental audio should remain stable, local audio effects are transient, while singing voice and music need to stay in tune \citep{zhang2024stylesinger,li2024robust,zhang2025versatile}.

To obtain caption data at this granularity, we construct \textbf{SwanData-Caption}, a pipeline that converts diverse speech-centered media audio into multi-level, multi-style captions. Its coverage design stage combines internal data with targeted synthetic subsets, so special styles, like pronunciation-challenging text, are not left to incidental coverage. The SwanData-Speech preprocessing stage provides each caption with reliable speech spans and speaker-attributed text anchors. The caption annotation stage then labels the environment, speakers, and fine-grained content fields, while the style-persona library substantially enriches the descriptions and characteristics of speaker styles. Finally, to improve data quality, the data refinement stage uses waveform filtering and caption auditing to select and polish high-quality, expressive, and accurately annotated data.

On the modeling side, we propose \textbf{SwanTale}, a multi-speaker expressive TTS and audio generation model that supports both zero-shot and instruct tasks. To handle multi-audio-modality generation, we design SwanVAE, whose architecture improves reconstruction quality while reducing the learning burden of the DiT prior. In the flow-based Transformer, we apply reward-conditioned quality control, which conditions inference on the highest quality without reinforcement learning, and Engram conditioning. To support both tasks and improve compatibility across audio modalities, we design Unified MoE with a task router and an audio router. We also use curriculum learning, moving from zero-shot ability to caption-conditioned generation, then to full-mixture training and expressive high-quality supervised fine-tuning. Finally, GRPO post-training improves pronunciation accuracy, generation stability, and caption-conditioned speaker-attribute control.

For experiments, we evaluate zero-shot monologue and dialogue TTS on SwanBench-Speech~\citep{pan2026swanbench}. We also evaluate instruction following on InstructTTSEval~\citep{huang2025instructttseval}, and acoustic quality on SwanBench-Scene. We also build SwanBench-Caption for heterogeneous instruct generation with dialogue speech and audio. Experimental results show that SwanTale leads on multiple key zero-shot and instruct metrics, achieves the best expressiveness scores in both tasks, and supports complex instruct generation involving multi-speaker speech and audio.

\section{Data Pipeline: SwanData-Caption}
\label{sec:data}

\begin{figure}[!htbp]
  \centering
  \includegraphics[width=\linewidth]{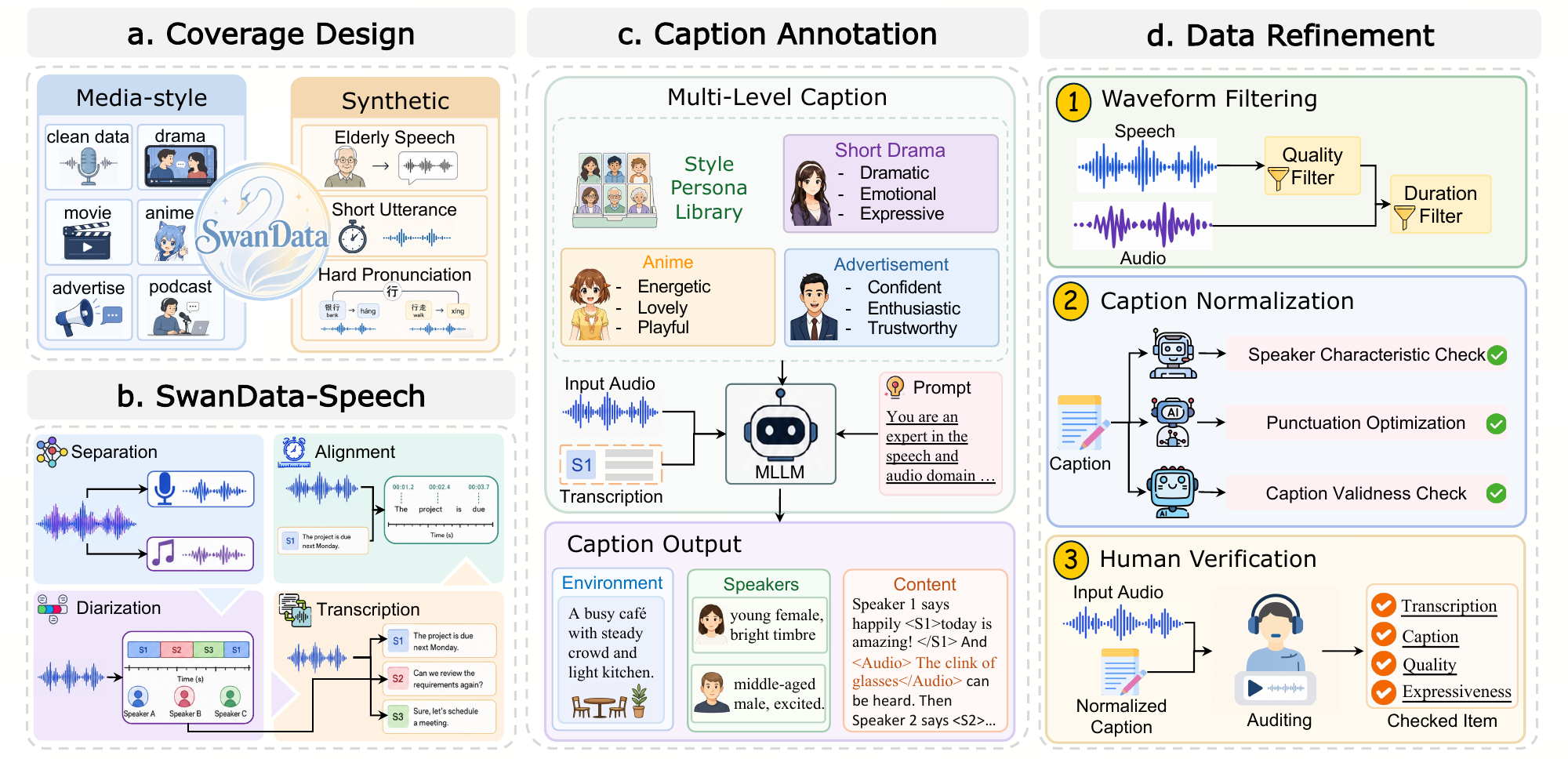}
  \caption{
  Overview of the four-stage SwanData-Caption data processing pipeline, including coverage design, SwanData-Speech preprocessing, caption annotation, and data refinement.}
  \label{fig:data_pipeline}
\end{figure}

SwanTale is trained from fine-grained captions rather than transcripts alone. For complex multi-speaker speech and audio data, basic speech transcription and speaker turns are not enough; the model also needs multi-level caption annotations~\citep{kim2019audiocaps}. Figure~\ref{fig:data_pipeline} summarizes our data processing pipeline. The following subsections correspond to the same four blocks: coverage design, SwanData-Speech preprocessing, caption annotation, and data refinement. The current mixture contains approximately 70M caption records.

\subsection{Coverage Design}

\noindent\textbf{Media-style coverage.}
SwanData-Caption is built from internal, real-world, speech-centered, and media-style data, covering speech, audio effects, and background music. Figure~\ref{fig:data_pipeline} shows representative examples of this coverage design, like short dramas, advertisements, and animations. The coverage spans diverse speaker densities, recording conditions, character styles, scene locations, sound-field conditions, persistent background effects, fine-grained delivery changes, and local audio effects. This design supports expressive speech generation in settings where speech and scene audio must be modeled together.

\noindent\textbf{Targeted synthetic coverage.}
Relying solely on real corpora leaves many rare scenarios underrepresented. Following SwanVoice~\citep{li2026swanvoice}, we incorporate three targeted synthetic subsets. A phoneme-aware TTS teacher~\citep{jiang2025megatts} is employed to generate these subsets to maintain pronunciation accuracy, with each containing 100k utterances.
(1) Elderly speech: Because elderly speech remains underrepresented in existing speech datasets~\citep{chen2025seniortalk}, we expand the limited elderly-speaker data to improve the model's demographic age coverage. The average duration is about 10 seconds.
(2) Short utterances in Chinese and English: These utterances range from single words and names to standard short sentences and have an average duration of 1.5 seconds. Many real-world TTS applications require sub-second spoken content, and this subset enhances model stability in such conditions.
(3) Challenging pronunciation targets: These are primarily Chinese texts featuring polyphonic characters, proper nouns, brand names, English insertions, and mixed-language phrases. For these pronunciation-sensitive examples, the model-facing target text remains unchanged, while the teacher model's input incorporates pronunciation hints to ensure the synthesized waveform reflects the intended reading. We process these data using pypinyin~\citep{mozillazg2023pypinyin}. Overall, human listeners are highly sensitive to errors in polyphonic characters and brand names, yet models encounter few such examples, and existing annotations are often inaccurate; therefore, this specific expansion is essential. The average duration is approximately 10 seconds.

\subsection{SwanData-Speech Preprocessing}

This stage reuses the SwanData-Speech backbone~\citep{li2026swanvoice} for speech-centered segments rather than pure audio-only clips. Its role is to prepare clean speech spans and reliable text anchors.

\noindent\textbf{Separation.}
For mixed-media audio, we first separate vocal and residual background streams with Ultimate Vocal Remover~\citep{ultimatevocalremovergui}. The vocal stream is used for higher-quality diarization, transcription, and alignment. We keep the original audio with the environmental sound field and local audio effects needed for captioning.

\noindent\textbf{Diarization.}
Speaker diarization is performed with the 3D-Speaker toolkit~\citep{chen20253d}, which combines VAD, CAM++ embeddings~\citep{wang2023campp}, and clustering. For single-speaker data, we keep segments between 1 and 60 seconds. For multi-speaker data, we allow segments up to 120 seconds and require at least two speaker turns whenever possible. Diarization model accuracy is limited, so we use it only for coarse data segmentation; real speaker discrimination is left to the subsequent caption annotations.

\noindent\textbf{Transcription.}
ASR is applied to the vocal stream with Seed-ASR 2.0~\citep{byteplus2026seedasr2}. We also use SenseVoice ASR~\citep{an2024funaudiollm} for an additional ASR-based pronunciation check. Specifically, we do not trust punctuation from these ASR tools, because their punctuation depends more on semantics than on actual pauses.

\noindent\textbf{Alignment.}
SwanAligner~\citep{li2026swanvoice} aligns the ASR transcript with the vocal stream and stores pause evidence for later post-processing. This module is related to forced-alignment and word-timestamping systems~\citep{rastorgueva2023nemo,bain2023whisperx,hu2025word}. Specifically, we let the subsequent caption annotation provide suitable punctuation according to semantics, such as exclamation marks and question marks, and then use the aligner results for correction, instead of making the aligner convert all punctuation into fixed commas and periods from the beginning.

\subsection{SwanData-Caption Annotation}

To provide structured supervision for controllable audio generation, we develop SwanData-Caption, an automatic annotation pipeline that converts raw audio and ASR transcripts into unified, fine-grained captions. 

\noindent\textbf{Annotation inputs.}
We use Seed2.0 Lite~\citep{bytedanceseed2026seed2} as the caption annotator. The annotator receives the target audio, the de-punctuated transcript, and a captioning prompt. In the transcript, ASR outputs from multiple speakers may be concatenated together, so the model is required to segment speakers, split sentences, and add semantically appropriate punctuation. The captioning prompt specifies a strict output format and strict behavioral requirements for the model, and provides suitable examples tailored to each dataset to be annotated. Our constraints include, for example, that Speaker IDs must be contiguous, every speaker span in Content must match an entry in Speakers, and the ASR text inside speaker tags must preserve the original language.

\noindent\textbf{Style-persona library.}
We find that, without sufficient examples, the model produces highly impoverished speaker styles, lacks plausible persona information, and often relies on only a few individual examples. Therefore, we design a style-persona library as a soft prior during annotation. In particular, we construct separate style matrices for three media families with strong stylistic priors: animation, short drama and film/TV drama, and advertisement/digital-human content. These matrices specify scene triggers, the preferred ordering of stable speaker descriptors, and the boundary between stable speaker profiles and transient local delivery. Animation uses a dubbing-style baseline with role-like vocal archetypes; short drama and film/TV drama allow plot-supported role identity and dramatic delivery; and advertisement/digital-human content emphasizes persona type, voice-style class, timbre, and stable product-pitch delivery. The library does not introduce labels that are not audible; instead, it guides the annotator to describe stable speaker traits in the Speakers field and leave transient emotion or emphasis to the Content field. The condensed style matrices are provided in Appendix~\ref{apx:caption_style_matrix}. For ordinary data, we omit these style matrices to avoid introducing unsupported attributes.

\noindent\textbf{Caption output.}
Each caption contains the three fields described in Table~\ref{tab:caption_metadata}. Scene location, sound-field impression, recording space, reverberation, and persistent background sounds or effects are assigned to Environment. Only speakers who actually speak are included in Speakers, where their detailed speaking styles are described. Fine-grained local style descriptions, such as emotion, together with local audio effects, are written in Content. The Content field supports both zero-shot and instruct TTS. An example caption is: 

{\color{seedblue}Environment: \{an antique courtyard interior with a quiet ambience; faint <Audio>wind and dripping water from the eaves</Audio> can be heard in the distance.\} Speakers: \{Speaker 1: a young woman, a reborn heroine, calm, with a low pitch and a cold tone.\} Content: \{Speaker 1 speaks in an icy and scrutinizing tone: <S1>What you owe me, I will make you spit it back out, mouthful by mouthful, with your own hands!</S1>\}.}

\begin{table}[!htbp]
  \centering
  \caption{Output-field definitions used in the caption annotation schema.}
  \label{tab:caption_metadata}
  \begin{tabular}{p{0.16\linewidth}p{0.74\linewidth}}
    \toprule
    Field & Description \\
    \midrule
    Environment & Scene-level environment and recording context, including location or place description, sound-field impression, room or recording-space cues, microphone characteristics, reverberation, background music, crowd murmur, traffic, wind, rain, electrical hum, keyboard tapping, appliance noise, distant footsteps, or other persistent background sound or effects that function as the scene bed. \\
    Speakers & The inventory of actually speaking subjects. Each speaker is described by perceived gender, age range, persona or role when audible from delivery, stable timbre, articulation, loudness tendency, speaking rate, accent, habitual style, and stable affective tendency. \\
    Content & A chronological content field and fine-grained local style description. Speech is wrapped by speaker tags such as \texttt{<S1>} and \texttt{</S1>}; local audio effects are wrapped by \texttt{<Audio>} and \texttt{</Audio>}. Fine-grained local style, including changes in emotion, volume, pace, pause, emphasis, hesitation, interruption, code-switching, and nearby effect context, is described around the tagged spans. \\
    \bottomrule
  \end{tabular}
\end{table}

\subsection{Data Refinement}

\noindent\textbf{Waveform filtering.}
To filter speech samples by acoustic quality, we apply the following waveform-level criteria only to speech data. Candidate speech clips are first scored with DNSMOS~\citep{reddy2021dnsmos} and reference-free estimates from torchaudio-SQUIM~\citep{kumar2023torchaudiosquim}, including STOI- and PESQ-related scores~\citep{zezario2020stoi,rix2001pesq}. In the default filtering configuration, speech samples are removed if PESQ is below 2.0, STOI is below 0.85, SI-SDR is below 0, or MOS is below 2.5. The duration filter keeps retained speech segments between 1 and 120 seconds, with an average duration of about 10 seconds across the retained speech set.

\noindent\textbf{Caption normalization.}
Caption normalization checks speaker characteristics, punctuation, and caption validity. SwanVerifier, a lightweight waveform-grounded attribute model described in Appendix~\ref{apx:swanverifier}, checks gender and age-range labels against the vocal stream and flags inconsistent speaker descriptions for removal. Punctuation is normalized only for spoken content. We use alignment data from SwanAligner to regularize pause marks and obvious boundary punctuation. If the original punctuation meets the pause requirements, we preserve it; otherwise, we replace or remove it. This allows special punctuation annotated by the annotator, such as exclamation marks and question marks, to be retained, rather than converting all punctuation into periods and commas. Caption-validity checks remove illegal speaker indices, unmatched speaker spans, missing required fields, unused speaker descriptions, or local text that disagrees with the transcript after punctuation stripping.

\noindent\textbf{Human verification.}
Human auditors review transcription accuracy, caption quality, audio quality, and expressiveness. They correct ASR errors, speaker-attribution mistakes, omitted local audio effects, inaccurate environment descriptions, and captions that over-interpret non-audible information. They also inspect failure modes that are often missed by objective metrics, like crowd leakage, separation artifacts, and speaker confusion in dense dialogue. In addition, they annotate audio-quality issues such as background noise, unclear articulation, severe word elision, unnatural accents, electronic artifacts, and audible editing traces.

Expressiveness is audited through group-wise best--worst comparison. Within each matched group and task type, we sample four valid candidates and ask annotators to select the most expressive and the least expressive one, considering naturalness, emotion strength, prosodic variation, and contextual appropriateness. 
Compared with MOS, this protocol does not require annotators to maintain a globally consistent absolute scoring standard across different speakers, contents, and tasks. Annotators only need to judge relative extremes within a controlled group, which reduces scale bias and calibration noise. Compared with pairwise A/B testing, our best--worst comparison method is much more annotation-efficient \citep{kiritchenko2017bestworst}.

\section{Method: SwanTale}
\label{sec:method}

This section introduces SwanTale. We first describe SwanVAE, the 48~kHz waveform-latent autoencoder. Then we present the flow-based Transformer, including content, caption, speaker-turn, Engram conditioning, and reward-conditioned quality control. We next introduce Unified MoE for zero-shot and instruct tasks with multiple audio modalities, followed by curriculum training, GRPO post-training, and inference procedure.

\begin{figure}[!htbp]
  \centering
    \includegraphics[width=1.0\textwidth]{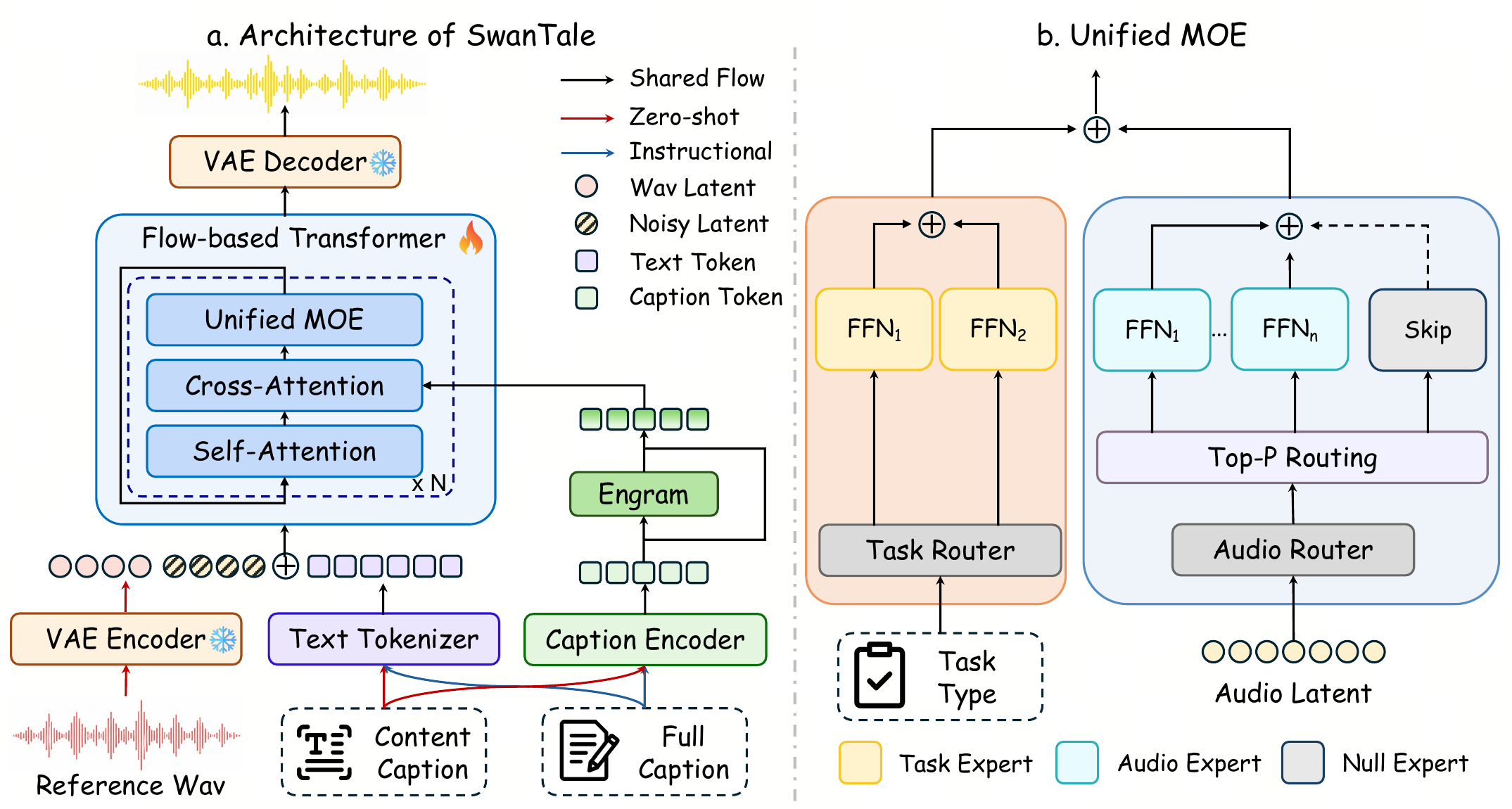}
  \caption{Overview of SwanTale. Figure (a) shows the architecture of SwanTale, and Figure (b) shows Unified MoE. 
  In (a), the zero-shot path supplies reference audio, while both tasks share text and caption. In (b), a task router selects experts at the sample level, while an audio router applies Top-$P$ routing over frame-level audio and null experts.
  }
  \label{fig:swantale_overview}
\end{figure}

\subsection{SwanVAE}

\begin{figure}[!t]
  \centering
  \includegraphics[
    width=1.0 \linewidth,
    pagebox=cropbox
  ]{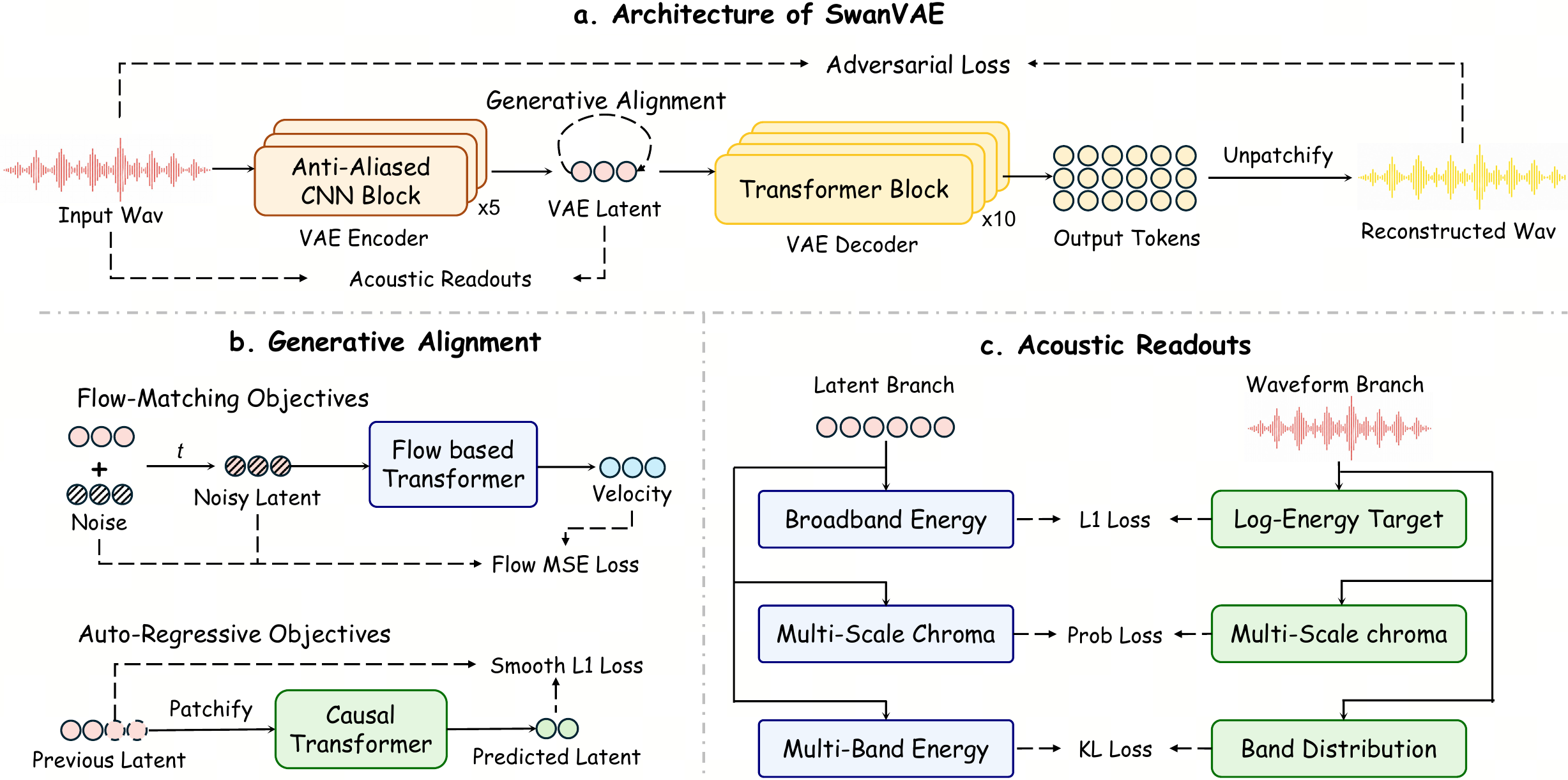}
  \caption{Overview of SwanVAE. (a) The anti-aliased convolutional encoder, Gaussian variational bottleneck, and local Transformer decoder. (b) Generative alignment through flow matching and causal latent prediction. (c) Energy, multi-scale chroma, and multi-band energy readouts with waveform-derived targets. The decoder receives posterior samples $\mathbf{z}$, while the alignment objectives operate on the posterior mean $\boldsymbol{\mu}_{\phi}$ exclusively during the SwanVAE training stage and do not enter the downstream generator at inference time. }
  \label{fig:swanvae_overview}
\end{figure}

SwanTale operates on the continuous acoustic latents produced by SwanVAE. Designing this latent space involves a three-way balance among reconstruction fidelity, representation compactness, and learnability by the downstream flow model. A lower latent rate shortens the sequence used for long-form generation, but it also requires each latent frame to carry more acoustic information. Encoding fine waveform detail improves reconstruction, yet the resulting latent distribution can become harder for the flow model to predict. SwanVAE represents 48~kHz mono audio as 96-dimensional latents at 25~Hz. The encoding path uses a local anti-aliased convolutional encoder and a Gaussian VAE bottleneck, and most waveform synthesis capacity is placed in a decoder-side Transformer Resampling Block adapted from SAME~\citep{parker2026same}. SwanVAE is intentionally restricted to local acoustic modeling, with longer-range sequence modeling handled by the downstream DiT. Figure~\ref{fig:swanvae_overview} summarizes the SwanVAE architecture and the training-only objectives used to shape its posterior mean.

\subsubsection{Architecture}

\noindent\textbf{Encoder.}
The encoder begins with a weight-normalized 1-D convolution, followed by five downsampling stages with rates $[4,4,4,5,6]$. Their product gives the 1920-sample hop required for a 25~Hz latent sequence. Each stage contains three residual units with dilation factors $1$, $3$, and $9$, followed by a strided projection. A fixed low-pass filter is applied before temporal decimation to reduce aliasing at large strides~\citep{zhang2019making}. A learnable high-frequency envelope shortcut with a small initial gain is added alongside the filtered path. The channel width grows from 64 to 1536. No Transformer layer is used in the encoder, whose waveform receptive field is approximately 0.95~s in the exact configuration used here.

\noindent\textbf{Variational bottleneck.}
Two $1\times1$ convolutional heads map the encoder output to the posterior mean $\boldsymbol{\mu}_{\phi}$ and log-variance $\log\boldsymbol{\sigma}_{\phi}^{2}$. For an input waveform $\mathbf{x}$, the posterior and its reparameterized sample are
\begin{equation}
q_{\phi}(\mathbf{z}\mid\mathbf{x})
=
\mathcal{N}\!\left(
\boldsymbol{\mu}_{\phi}(\mathbf{x}),
\operatorname{diag}\!\left(\boldsymbol{\sigma}_{\phi}^{2}(\mathbf{x})\right)
\right),
\qquad
\mathbf{z}
=
\boldsymbol{\mu}_{\phi}(\mathbf{x})
+
\boldsymbol{\sigma}_{\phi}(\mathbf{x})\odot\boldsymbol{\epsilon},
\quad
\boldsymbol{\epsilon}\sim\mathcal{N}(\mathbf{0},\mathbf{I}).
\end{equation}
The sampled latent $\mathbf{z}$ is passed to the decoder during VAE training. SwanTale uses the globally normalized posterior mean $\boldsymbol{\mu}_{\phi}$ as its deterministic acoustic representation.

\noindent\textbf{Decoder.}
The decoder uses the decoder-side Transformer Resampling Block (TRB) introduced in SAME~\citep{parker2026same}. Each latent vector is projected to the decoder width and interleaved with six learnable output tokens. A local bidirectional Transformer processes the combined sequence, after which each output token is projected to a 320-sample waveform patch. The six output patches associated with one latent frame cover $6\times320=1920$ samples, or 40~ms of audio. Self-attention couples neighboring latent and output tokens before projection, so adjacent waveform patches are reconstructed with shared context even though they are concatenated without overlap at the final waveform output during both training and inference.

The asymmetric allocation reflects the role of each path. The encoder determines the temporal support and statistics of the latent target seen by SwanTale, so every reduction stage retains an explicit low-pass operation and the encoding context stays local. Once the latent sequence has been formed, the decoder can mix nearby frames and spend more capacity on waveform synthesis. Its local context supports phase continuity, patch boundaries, transient structure, and high-frequency detail.

\noindent\textbf{Receptive field design.} We deliberately keep the temporal context of SwanVAE bounded. The convolutional encoder ties each latent to a local waveform neighborhood, while the decoder uses local attention to coordinate waveform details across adjacent patches. With the current configuration, the encoder covers approximately 0.95 seconds of waveform, and the end-to-end theoretical dependency span is about 3.23 seconds. The downstream DiT subsequently operates over the full latent sequence to model dependencies beyond this local acoustic context across the complete generated sequence.

\subsubsection{Reconstruction and Adversarial Training}

We train SwanVAE on fixed-duration waveform segments sampled at random from the training recordings. This exposes the model to different local regions of long recordings while keeping the training context consistent with the local role of SwanVAE. Similar to the on-the-fly source mixing used in EnCodec~\citep{defossez2022encodec}, we randomly mix pairs of segments and use the mixture as both the encoder input and reconstruction target. This covers overlapping acoustic sources, and we also observed smoother decoded transitions under latent interpolation.

SwanVAE reconstructs each training waveform $\mathbf{x}$ from the reparameterized posterior sample $\mathbf{z}$. The reconstruction loss combines a multi-resolution complex STFT loss, a multi-resolution multi-band Mel loss, and a frame-wise energy loss, as summarized by the following objective:
\begin{equation}
\mathcal{L}_{\mathrm{rec}}
=
\lambda_{\mathrm{stft}}\mathcal{L}_{\mathrm{stft}}
+
\lambda_{\mathrm{mel}}\mathcal{L}_{\mathrm{mel}}
+
\lambda_{\mathrm{eng}}\mathcal{L}_{\mathrm{eng}}.
\end{equation}
The complex STFT loss compares spectral magnitude and phase across multiple time-frequency resolutions. The multi-band Mel loss provides spectral-envelope supervision over different frequency ranges and analysis windows, while $\mathcal{L}_{\mathrm{eng}}$ matches frame-wise log energy on the 40-ms latent grid. We regularize the posterior toward a unit Gaussian with a KL penalty, using $\lambda_{\mathrm{KL}}=0.02$.

Fine waveform structure remains underconstrained by these regression losses. We therefore train three discriminator families: a multi-period discriminator (MPD)~\citep{kong2020hifi}, a multi-resolution discriminator (MRD)~\citep{jang2021univnet}, and a multi-band complex STFT discriminator (MBCSD)~\citep{kumar2023dac}. MPD is sensitive to periodic waveform patterns, while MRD evaluates time-frequency structure at several STFT resolutions. MBCSD operates on the real and imaginary components of complex STFTs. At each resolution, it partitions the frequency axis into fixed bands and processes each band with a separate convolutional stack. This provides band-specific discrimination paths over the spectrum up to 24~kHz during training.

Let $\mathcal{D}=\{\mathrm{MPD},\mathrm{MRD},\mathrm{MBCSD}\}$ denote the discriminator families. During training, we optimize the generator with the following waveform objective:
\begin{equation}
\mathcal{L}_{\mathrm{wav}}
=
\mathcal{L}_{\mathrm{rec}}
+
\lambda_{\mathrm{KL}}\mathcal{L}_{\mathrm{KL}}
+
\lambda_{\mathrm{adv}}
\sum_{D\in\mathcal{D}}w_D\mathcal{L}_{\mathrm{adv}}^{D}
+
\lambda_{\mathrm{fm}}
\sum_{D\in\mathcal{D}}\mathcal{L}_{\mathrm{fm}}^{D},
\end{equation}
where $w_D$ controls the adversarial contribution of each discriminator family. Feature matching is computed from their intermediate activations for real and reconstructed audio. All discriminators are discarded after training and add no cost to SwanVAE decoding at inference time.

\subsubsection{Latent Alignment Objectives}

Waveform objectives constrain the reconstructed signal but leave considerable freedom in how acoustic information is arranged in the posterior mean. At a latent rate of 25~Hz, fine waveform detail may be carried by sharp changes between adjacent frames, increasing the burden on the downstream flow model. Related work on image and video autoencoders has also associated excessively high-frequency latent components with more difficult diffusion modeling~\citep{skorokhodov2025diffusability}. SwanVAE therefore applies several weak alignment objectives to the posterior mean $\boldsymbol{\mu}_{\phi}$ during training as auxiliary guidance.

\noindent\textbf{Generative alignment.}
Following SAME~\citep{parker2026same}, we jointly train a lightweight unconditional predictor on the latent sequence using the standard flow-matching objective. The predictor is optimized with its full training loss, while the gradient passed from this objective to the encoder is scaled separately and kept small. It provides a direct signal about how readily the current latent distribution can be modeled by a flow network, without allowing the auxiliary predictor to dominate waveform reconstruction.

We also train a causal predictor to estimate future latent patches from their preceding context. Future prediction is well established in speech representation learning~\citep{oord2018representation,chung2019apc}, and latent-domain predictive coding has also been used to remove temporal redundancy in neural speech codecs~\citep{jiang2023latent}. In SwanVAE, the predictor operates directly on continuous posterior-mean patches and provides only a weak target-side gradient to the encoder. The prediction residual therefore measures the part of local latent evolution that cannot be inferred from recent history. Compared with a fixed temporal-difference penalty, the learned predictor can accommodate locally predictable changes while discouraging abrupt and weakly structured variation. We keep the encoder-facing gradient small so that the global latent geometry remains governed primarily by reconstruction and KL regularization throughout joint SwanVAE training.

\noindent\textbf{Semantic and acoustic readouts.}
Following the semantic regression objective in SAME~\citep{parker2026same}, lightweight regressors predict octave-specific, multi-scale chroma distributions from each latent frame. Here, semantic alignment refers to perceptually meaningful structure within a local audio context, including pitch-class and harmonic organization represented at the latent-frame scale.

Two additional readouts predict normalized frame energy and the relative energy distribution across frequency bands from each latent frame. To increase variation in effective bandwidth, we occasionally downsample a 48~kHz training crop to a lower sampling rate and resample it back to 48~kHz. This keeps the waveform interface fixed while exposing the encoder to signals with different usable spectral extents. The multi-band energy readout encourages the latent representation to retain these differences. For example, a 48~kHz recording derived from 24~kHz audio typically contains little energy above 12~kHz. Since the band-energy target is normalized across frequency bands, it describes spectral allocation independently of overall loudness.

The reconstruction decoder continues to receive the reparameterized sample $\mathbf{z}$, whereas the alignment objectives act on the posterior mean $\boldsymbol{\mu}_{\phi}$. After training, SwanTale uses the globally normalized posterior mean as its deterministic acoustic target. All auxiliary predictors and readout heads are discarded.

\subsection{Flow-based Transformer}

Following SwanVoice \citep{li2026swanvoice}, SwanTale trains a non-causal diffusion Transformer with flow matching as the backbone to avoid repetition and other instabilities common in autoregressive systems, while preserving the integrity of local audio effects and environmental soundscapes. Specifically, SwanTale maps the inputs of the two tasks to a latent audio trajectory, as shown in Figure~\ref{fig:swantale_overview}(a). For these two tasks, zero-shot samples use a content caption to control local style and reference audio to control global acoustic information, while instruct samples use a full caption to control all information. The denoiser is non-causal because these controls interact across the whole audio sequence. To realize this interface, SwanTale combines caption, text, and speaker conditioning for semantic alignment, reward-conditioned quality control for global acoustic preference, Engram conditioning for recurring caption patterns, and a flow-matching DiT backbone for conditioned generation.

\noindent\textbf{Caption, text, and speaker conditioning.}
General instruct-generation systems often encode captions and text with a shared tokenizer and rely on an LM backbone, typically inherited from a pretrained LLM \citep{yang2025qwen3} with strong language-understanding ability, for instruction understanding and generation \citep{zhang2025mimoaudio,qwen2026qwen3tts}. Since SwanTale uses a DiT backbone, we need to preserve lexical accuracy while strengthening caption understanding without overloading the model. Therefore, the conditioning stack separates caption-level control from text alignment.

The caption branch uses a Qwen-family text encoder \citep{bai2023qwen} to understand the caption and encode it into embeddings, which are injected into all DiT layers through cross-attention. In addition, we add a set of label embeddings aligned to the caption-embedding length. Different embeddings are used to distinguish speech content, local audio effect descriptions, environment information, and other descriptions, providing acoustic prior information and reducing the learning burden on cross-attention.

The spoken content is tokenized with the CosyVoice 2.0 tokenizer \citep{du2024cosyvoice}, and the resulting tokens are processed by a lightweight Transformer text encoder. Unlike the filler-token expansion adopted in SwanVoice \citep{li2026swanvoice}, SwanTale length-normalizes this branch by interpolating text-encoder hidden states onto the audio-latent timeline before concatenation with the noised latent stream. This allows normal training and generation even when the boundary text is longer than the latent sequence. Speaker-turn embeddings are derived from the structured speaker tags, aligned with the text embeddings, and injected into the text path.

\noindent\textbf{Reward-conditioned quality control.}
During preprocessing, SwanData-Caption annotates quality scores for the STOI-, PESQ-, SI-SDR-, and MOS-related metrics described in Section~\ref{sec:data}. The data pipeline filters out very low-quality samples, but the remaining data still spans different quality levels, and we want generated speech to be as high-quality as possible. Therefore, we add an explicit quality caption and quality flag so the model can recognize data quality and use it as a controllable signal.
When all four scores are available, SwanTale converts them into a short quality caption and appends it to the global caption before the content field. Specifically, the implemented quality caption template is: {\small\ttfamily Quality: speech clarity \{STOI level\}; noise level \{SI-SDR level\}; signal naturalness \{PESQ level\}; listening quality \{MOS level\}.}

The quality scores are also mapped into the quality flag $q\in\{\mathrm{low},\mathrm{normal},\mathrm{high},\mathrm{unknown}\}$. During training, this flag can be replaced by the unknown value under dropout to support classifier-free guidance; during inference, we use the high-quality caption and flag to bias generation toward clearer and more natural speech.

This makes SwanTale a reward-conditioned policy~\citep{kumar2019rewardconditioned}. The four waveform-quality scores act as a reward that is supplied as part of the condition rather than optimized against, so the model learns how each quality level is acoustically realized; at inference, the reward is fixed to its maximum, and generation is always requested at the highest quality. Compared with explicitly optimizing a quality reward, this requires no rollout, no reward model in the loop, and no additional sampling. It also keeps every retained sample useful for training: moderate-quality data still contributes to coverage of speakers, scenes, and audio effects, with its quality level marked rather than discarded, preserving its contribution across the full data mixture.

\noindent\textbf{Engram conditioning.}
In real captions, many patterns recur across samples with stable and distinctive meanings, such as persona descriptions like an energetic girl, or common audio effects like a train whistle. Pure attention can learn these patterns, but it must also handle long-range acoustic planning. To reduce this burden and make fixed caption patterns recognizable even in long captions, SwanTale adds an Engram memory layer \citep{cheng2026conditionalmemory} to the caption branch. Engram separates fixed-pattern recognition from broader acoustic planning without introducing another full language encoder. Concretely, after the caption encoder output is projected, it is also processed by Engram and added back as a memory update. The resulting caption representation is then used as the cross-attention context.

For a caption token sequence $\mathbf{c}=(c_1,\ldots,c_L)$ and an order set $\mathcal{N}$, the $n$-gram window centered at position $i$ is $w^{(n)}_i=c_{\,i-\lfloor (n-1)/2\rfloor\,:\,i+\lfloor n/2\rfloor}.$
The original formulation uses suffix windows because an autoregressive backbone cannot see its right context; the caption branch here is non-causal, so we center the window instead. The memory hashes each window into $K$ head-specific tables and concatenates every retrieved slot:
\begin{equation}
\mathbf{e}_i
=
\operatorname{concat}
\left(
\left\{
\operatorname{Engram}_{n,k}
\left(\operatorname{hash}_{k}(w^{(n)}_i)\right)
\right\}_{n\in\mathcal{N},\,k=1,\ldots,K}
\right).
\end{equation}
We set $\mathcal{N}=\{2,3\}$. Concatenating before read-out lets a single pair of projections weigh the orders against each other, rather than forcing every order to contribute equally.
Given the projected caption embedding $\mathbf{u}_i$, SwanTale injects this memory by a gated residual update:
\begin{equation}
\tilde{\mathbf{u}}_i
=
\mathbf{u}_i
+
\sigma\!\left(
\frac{\operatorname{RMSNorm}(\mathbf{u}_i)^{\top}\operatorname{RMSNorm}(W_K\mathbf{e}_i)}{\sqrt{d}}
+
b
\right)
W_V\mathbf{e}_i.
\end{equation}

Here, $d$ is the model dimension. The gate is instantiated with two branches that share the tables and $W_V$ while keeping separate $W_K$, and their gated outputs are averaged. The learnable bias $b$ is initialized to a negative value, so the memory path starts nearly closed and gradually opens during training. The dot product makes the gate content-dependent within each caption, allowing the model to use Engram more strongly for structured markers and less strongly for free-form natural language. In this way, the memory becomes part of the caption-conditioning stack and strengthens the model's understanding of fixed-pattern phrases.

\noindent\textbf{Flow-matching DiT backbone.}
SwanTale builds on a non-causal flow-matching DiT architecture \citep{li2026swanvoice} with three additions: Engram-enhanced caption cross-attention, reward-conditioned quality control, and Unified MoE feed-forward layers. Each block contains self-attention over the latent timeline, cross-attention over the Engram-enhanced caption tokens, and a Unified MoE feed-forward branch. Timestep embeddings modulate the block through AdaLN-Zero adapters \citep{peebles2023scalable}, and RMSNorm is used for stable deep Transformer optimization \citep{zhang2019root}. The quality-flag embedding is added to the same global conditioning stream as the timestep embedding.

SwanTale uses one backbone for joint instruct and zero-shot training. Let $\tau\in\{\mathrm{inst},\mathrm{zero}\}$ denote the task type and let $\mathbf{x}^{\star}\in\mathbb{R}^{T\times d_z}$ denote the target latent sequence produced by the frozen SwanVAE. The two tasks differ only in their respective caption inputs and context masks:
\begin{equation}
\mathbf{c}^{(\tau)}
=
\begin{cases}
\mathbf{c}_{\mathrm{full}}, & \tau=\mathrm{inst},\\
\mathbf{c}_{\mathrm{content}}, & \tau=\mathrm{zero},
\end{cases}
\qquad
\mathbf{m}^{(\tau)}
=
\begin{cases}
\mathbf{0}, & \tau=\mathrm{inst},\\
\mathbf{m}_{\mathrm{prompt}}, & \tau=\mathrm{zero},
\end{cases}
\qquad
\mathbf{r}^{(\tau)}=\mathbf{m}^{(\tau)}\odot\mathbf{x}^{\star}.
\end{equation}
Here $\mathbf{m}^{(\tau)}$ is a context mask: instruct samples have no prompt context and therefore use the whole latent sequence as the generation target, whereas zero-shot samples use prompt frames as reference context and generate only the remaining frames. Following recent non-AR speech generation systems \citep{lipman2022flow,chen2024f5}, we sample Gaussian noise $\boldsymbol{\epsilon}\sim\mathcal{N}(\mathbf{0},\mathbf{I})$ and a time $t\sim\mathcal{U}(0,1)$, then noise only the task-specific generation region:
\begin{equation}
\tilde{\mathbf{x}}^{(\tau)}_t
=
\left(1-\mathbf{m}^{(\tau)}\right)
\odot
\left((1-t)\boldsymbol{\epsilon}+t\mathbf{x}^{\star}\right).
\end{equation}
The DiT receives $\tilde{\mathbf{x}}^{(\tau)}_t$, the context latent $\mathbf{r}^{(\tau)}$, and a learned context-mask embedding that distinguishes generated frames from reference frames. It predicts a velocity field conditioned on the content tokens $\mathbf{y}$, caption $\mathbf{c}^{(\tau)}$, speaker turns, and task-specific reference context:
\begin{equation}
\hat{\mathbf{v}}_{\theta}
=
f_{\theta}
\left(
\tilde{\mathbf{x}}^{(\tau)}_t,
t,
\mathbf{y},
\mathbf{c}^{(\tau)},
\mathbf{r}^{(\tau)},
\mathbf{m}^{(\tau)}
\right).
\end{equation}
The training loss is a masked mean-squared error over the task-specific generation region:
\begin{equation}
\label{eq:swantale_flow_loss}
\mathcal{L}_{\mathrm{flow}}
=
\mathbb{E}
\left[
\frac{
\left\|\left(1-\mathbf{m}^{(\tau)}\right)\odot
\left(\hat{\mathbf{v}}_{\theta}-(\mathbf{x}^{\star}-\boldsymbol{\epsilon})\right)
\right\|_2^2}
{\max\left(1,\sum_{i=1}^{T}\left(1-\mathbf{m}^{(\tau)}_i\right)\right)}
\right].
\end{equation}
Under this formulation, instruct samples provide full-caption supervision over the entire latent trajectory, while zero-shot samples provide content-caption supervision with reference audio excluded from the loss but available as context. The same velocity parameterization and objective therefore train a unified backbone for both instruction-following generation and prompt-conditioned zero-shot generation.

\subsection{Unified MoE}

SwanTale handles both instruct and zero-shot tasks within a single network, covering multi-speaker expressive speech, general audio, occasional singing voice, and music. Speech regions must preserve content, prosody, and speaker continuity while modeling speaker-dependent expression. General audio spans two regimes: environmental sound forms smooth and persistent structure, whereas local audio effects are transient and require sharply localized acoustic shaping. Singing voice adds sustained pitch contours, and music carries melodic and harmonic regularity without lexical content. Processing all of them with the same dense feed-forward parameters forces these heterogeneous acoustic patterns to compete for one shared set of weights.

We therefore introduce Unified MoE in our flow-based Transformer, a caption-conditioned dynamic-capacity sparse feed-forward module, as illustrated in Figure~\ref{fig:swantale_overview}(b). It dynamically allocates model capacity according to the generation task, the current acoustic state, and the diffusion time, providing specialized transformations for complex regions while controlling the overall computation. Unified MoE builds on sparse-expert Transformers \citep{fedus2022switch,zoph2022stmoe}, DeepSeek-style expert specialization \citep{dai2024deepseekmoe}, and auxiliary-loss-free routing bias \citep{wang2024auxiliarylossfree}. Its routing mechanism operates at two levels. A task router selects sample-level shared experts that capture stable priors for instruct and zero-shot generation, while an audio router applies dynamic Top-$P$ routing to each latent-frame hidden state to model frame-level acoustic variation. The hidden states received by the audio router have already passed through self-attention and caption cross-attention and therefore incorporate information from the caption, text stream, speaker turns, and reference audio.

\noindent\textbf{Expert layout.}
Every second DiT feed-forward layer is replaced with a Dynamic Top-$P$ MoE-FFN layer. The dense layers retained between sparse layers provide a stable shared path, while inserted MoE layers add capacity for heterogeneous acoustic patterns. Each MoE-FFN contains $R$ routed audio experts, $S$ task-shared experts, and $U$ null experts in every sparse layer.

The three expert types operate at different levels of granularity. Task-shared experts encode priors that remain stable throughout a sample: instruct generation places greater emphasis on caption following and the composition of multiple audio events, whereas zero-shot generation relies more strongly on prompt-speaker preservation. Routed audio experts provide input-dependent frame-level specialization for speaker changes, overlapping speech, expressive variation, local audio effects, and complex background textures. Null experts act as skip paths without an additional feed-forward transformation, reducing expert computation for stable backgrounds, near-silent regions, and other frames that do not require frame-level specialization.

Let $\tau\in\{\mathrm{inst},\mathrm{zero}\}$ denote the task type. The task router selects a set of shared experts $\mathcal{T}_{\tau}$ that is reused across all sparse layers and frames of each sample. The shared branch is defined as
\begin{equation}
\mathbf{o}_{\mathrm{shared}}(\mathbf{h},\tau)
=
\sum_{j\in\mathcal{T}_{\tau}}
E_j^{\mathrm{shared}}(\mathbf{h}).
\end{equation}
This task-routing branch injects sample-level task priors, while the audio-routing branch is designed to model frame-level acoustic variation over the latent sequence.

\noindent\textbf{Latent- and time-conditioned audio routing.}
At diffusion time $t$, let $\mathbf{h}_{\ell,m}$ denote the hidden state of frame $m$ after self-attention and caption cross-attention in the $\ell$-th DiT block. The DiT time embedding $\mathbf{e}_t$ is linearly projected and added to the corresponding frame representation:
\begin{equation}
\mathbf{r}_{\ell,m}
=
\mathbf{h}_{\ell,m}
+
W_t\mathbf{e}_t.
\end{equation}
The audio router computes the base router logits from $\mathbf{r}_{\ell,m}$:
\begin{equation}
\boldsymbol{\ell}_{\ell,m}
=
W_g\mathbf{r}_{\ell,m}
+
\mathbf{b}^{\mathrm{null}}(t),
\end{equation}
where $W_g$ projects the frame representation into the candidate space of routed audio experts and null experts. We then compute the expert-selection logits as follows:
\begin{equation}
\mathbf{a}_{\ell,m}
=
\boldsymbol{\ell}_{\ell,m}
+
\mathbf{b},
\end{equation}
where $\mathbf{b}$ is a load-correction bias for routed experts and is zero on null-expert dimensions. The expert set is determined from $\mathbf{a}_{\ell,m}$, while the combination weights within the selected set are computed from $\boldsymbol{\ell}_{\ell,m}$ \citep{wang2024auxiliarylossfree}.

For routed expert $i$, let $f_i$ denote its fraction of recent non-null assignments among the routed experts, with a target average load of $1/R$. We use the following continuous and clipped bias update:
\begin{equation}
b_i
\leftarrow
\operatorname{clip}
\left(
b_i
+
\eta
\left(
\frac{1}{R}-f_i
\right),
-B,B
\right).
\end{equation}
The selection score of an overloaded expert is reduced, whereas an underutilized expert receives a higher score. If a layer produces no non-null assignments, its routing bias is left unchanged.

\noindent\textbf{Time-aware expert budget.}
The required amount of expert computation varies with diffusion time. Different denoising stages emphasize global structure, speaker and event arrangement, timbre, and local acoustic detail to different degrees. We predict a learned time-dependent budget from the time embedding:
\begin{equation}
q(t)
=
\sigma(W_b\mathbf{e}_t).
\end{equation}
This budget jointly controls the Top-$P$ threshold, the null-expert bias, and the expert capacity:
\begin{align}
p(t)
&=
p_{\min}
+
(p_{\max}-p_{\min})q(t),\\
b_{\mathrm{null}}(t)
&=
b_{\max}^{\mathrm{null}}
+
\left(
b_{\min}^{\mathrm{null}}
-
b_{\max}^{\mathrm{null}}
\right)q(t),\\
c(t)
&=
c_{\min}
+
(c_{\max}-c_{\min})q(t).
\end{align}
The vector $\mathbf{b}^{\mathrm{null}}(t)$ takes $b_{\mathrm{null}}(t)$ on the null-expert dimensions and zero on the routed-expert dimensions. A larger $q(t)$ yields a higher cumulative-probability threshold, a weaker preference for null routing, and greater expert capacity. A smaller $q(t)$ reduces additional expert transformations and shifts more computation toward task-shared and null paths at the same denoising stage.

\noindent\textbf{Dynamic Top-$P$ routing with annealed Gumbel mixing.}
The number of experts required varies substantially across frames. Stable backgrounds and near-silent regions typically require few additional transformations, whereas speaker changes, environmental sound fields, and superimposed global audio events may benefit from multiple experts. Fixed Top-$K$ routing assigns the same number of experts to every frame and cannot adapt to this variation. Unified MoE therefore uses dynamic Top-$P$ routing.

During training, we first draw Gumbel noise for all routed audio and null experts:
\begin{equation}
g_{\ell,m, i}
=
-\log
\left(
-\log u_{\ell,m,i}
\right),
\qquad
u_{\ell,m, i}
\sim
\mathcal{U}(0,1).
\end{equation}
The Gumbel-Softmax selection distribution is then computed from the bias-adjusted selection logits:
\begin{equation}
\pi^{\mathrm{g,sel}}_{\ell,m,i}
=
\frac{
\exp
\left(
\frac{a_{\ell,m,i}+g_{\ell,m,i}}{\tau_g(n)}
\right)
}{
\sum_{j=1}^{R+U}
\exp
\left(
\frac{a_{\ell,m,j}+g_{\ell,m,j}}{\tau_g(n)}
\right)
}.
\end{equation}
For each frame, candidate experts are sorted in descending order of $\boldsymbol{\pi}^{\mathrm{g,sel}}_{\ell,m}$, and the smallest prefix whose cumulative probability reaches $p(t)$ is selected as $\mathcal{S}_{\ell,m}$.

The Gumbel temperature $\tau_g(n)$ is gradually annealed over optimization step $n$:
\begin{equation}
\tau_g(n)
\searrow
\tau_{\min}
>
0.
\end{equation}
Following the Gumbel-Softmax annealing strategy \citep{jang2016categorical}, a higher temperature at the beginning of training produces smoother routing distributions and encourages exploration over more expert combinations. Gradually, as the temperature decreases, the distribution becomes sharper, and the experts develop distinct roles, reducing premature concentration on a single expert early in optimization.

Using the same Gumbel perturbations, the mixture weights are computed from the base router logits and then renormalized over the selected expert set:
\begin{equation}
\tilde{\pi}_{\ell,m,i}
=
\frac{
\exp
\left(
\frac{\ell_{\ell,m,i}+g_{\ell,m,i}}{\tau_g(n)}
\right)
}{
\sum_{j\in\mathcal{S}_{\ell,m}}
\exp
\left(
\frac{\ell_{\ell,m,j}+g_{\ell,m,j}}{\tau_g(n)}
\right)
},
\qquad
i\in\mathcal{S}_{\ell,m}.
\end{equation}

At inference time, the Gumbel noise is removed. The deterministic selection distribution is
\begin{equation}
\boldsymbol{\pi}^{\mathrm{sel}}_{\ell,m}
=
\operatorname{softmax}
\left(
\mathbf{a}_{\ell,m}
\right).
\end{equation}
Candidate experts are sorted according to $\boldsymbol{\pi}^{\mathrm{sel}}_{\ell,m}$, and the smallest prefix whose cumulative probability reaches $p(t)$ is selected as $\mathcal{S}_{\ell,m}$. The corresponding mixture weights are
\begin{equation}
\bar{\pi}_{\ell,m,i}
=
\frac{
\exp(\ell_{\ell,m,i})
}{
\sum_{j\in\mathcal{S}_{\ell,m}}
\exp(\ell_{\ell,m,j})
},
\qquad
i\in\mathcal{S}_{\ell,m}.
\end{equation}
We use $\tilde{\pi}_{\ell,m,i}$ during training and $\bar{\pi}_{\ell,m,i}$ during inference, and denote both by $\hat{\pi}_{\ell,m,i}$ below. Let
\begin{equation}
\mathcal{S}_{\ell,m}^{\mathrm{r}}
=
\mathcal{S}_{\ell,m}
\cap
\{1,\ldots,R\}
\end{equation}
denote the selected routed audio experts. The audio-branch output is
\begin{equation}
\mathbf{o}_{\mathrm{audio}}
(\mathbf{h}_{\ell,m},t)
=
\sum_{i\in\mathcal{S}_{\ell,m}^{\mathrm{r}}}
\hat{\pi}_{\ell,m,i}
E_i^{\mathrm{audio}}
(\mathbf{h}_{\ell,m}).
\end{equation}
When $\mathcal{S}_{\ell,m}^{\mathrm{r}}=\varnothing$, the routed audio branch produces a zero output:
\begin{equation}
\mathbf{o}_{\mathrm{audio}}
(\mathbf{h}_{\ell,m},t)
=
\mathbf{0}.
\end{equation}
Null experts participate in the normalization over selected experts but produce no feed-forward output. When null and routed audio experts are selected together, the probability mass assigned to null experts reduces the magnitude of the audio-branch transformation. If only null experts are selected, the frame skips the routed audio branch while retaining the task-shared branch and the original residual path of the DiT block.

The final MoE-FFN output combines the shared and routed audio branches as follows:
\begin{equation}
\mathbf{o}_{\mathrm{MoE}}
(\mathbf{h}_{\ell,m},t,\tau)
=
\mathbf{o}_{\mathrm{shared}}
(\mathbf{h}_{\ell,m},\tau)
+
\mathbf{o}_{\mathrm{audio}}
(\mathbf{h}_{\ell,m},t),
\end{equation}
which is combined with the input through the original residual connection of the DiT block.

\noindent\textbf{Capacity and auxiliary objectives.}
Expert capacity limits the number of assignments received by each routed expert within a batch. Without capacity control, a large number of frames may be concentrated on a single expert, increasing peak memory usage and weakening functional specialization.

Let $M$ denote the total number of non-null frame-expert assignments produced by Top-$P$ routing before capacity truncation. The capacity of each routed expert is
\begin{equation}
\mathrm{cap}
=
\left\lceil
c(t)\frac{M}{R}
\right\rceil.
\end{equation}
If an expert receives more assignments than its capacity, only the assignments with the largest mixture weights are retained, and the remaining assignments are dropped as overflow. If all frames in the current layer select only null experts, then $M=0$ and routed-expert dispatch and capacity truncation are skipped. Expert dropout is further applied during training to reduce dependence on fixed expert combinations.

Meanwhile, we use lightweight auxiliary objectives for router-logit instability and null-routing collapse:
\begin{equation}
\mathcal{L}_{\mathrm{MoE}}
=
\lambda_z\mathcal{L}_z
+
\lambda_{\mathrm{null}}\mathcal{L}_{\mathrm{null}}.
\end{equation}
The router z-loss is applied directly to the base router logits during training:
\begin{equation}
\mathcal{L}_z
=
\mathbb{E}_{\ell,m}
\left[
\left(
\log
\sum_j
\exp(\ell_{\ell,m,j})
\right)^2
\right],
\end{equation}
which constrains the magnitude of the router logits and improves numerical stability \citep{zoph2022stmoe}. The null-collapse penalty $\mathcal{L}_{\mathrm{null}}$ is the average probability mass assigned to null experts by the base routing distribution, preventing the routed audio branch from remaining inactive throughout training.

The complete SwanTale training objective is defined as follows:
\begin{equation}
\label{eq:swantale_total_loss}
\mathcal{L}
=
\mathcal{L}_{\mathrm{flow}}
+
\omega_{\mathrm{MoE}}(n)
\mathcal{L}_{\mathrm{MoE}},
\end{equation}
where $\omega_{\mathrm{MoE}}(n)$ is linearly annealed during early training to a small nonzero floor. The stronger initial auxiliary signal stabilizes routing early in training, while later specialization is driven by the flow-matching objective.

Overall, Unified MoE combines a task-level shared path, frame-level dynamic expert routing, and a diffusion-time-aware computation budget to provide adaptive capacity for heterogeneous speech and scene audio within a unified model, without imposing the same fixed computation on every acoustic region.

\subsection{Curriculum Learning}

We use SwanVoice as the reference design for the zero-shot foundation stage when training the SwanTale base model. The model first learns speaker-conditioned speech generation from single-speaker and multi-speaker zero-shot data, and is then adapted to caption-conditioned instruct generation. This ordering is important because caption conditioning introduces longer contexts, text-described speaker attributes, environment descriptions, and local audio effects. Introducing all these conditions before the speech prior is sufficiently stable makes both alignment and conditional modeling more difficult \citep{bengio2009curriculum}.

\textbf{1) Zero-shot base model training.}
In the first stage, we train a zero-shot base model in the SwanVAE latent space. The training recipe follows the SwanVoice pretraining setup for single-speaker and multi-speaker speech, using only transcript text and optional reference audio as conditions. This preserves the reference-audio pathway required for zero-shot inference without making generation entirely dependent on a reference signal. We first train on internal single-speaker zero-shot data. We then introduce concatenated speech containing one to four speakers, together with real one- to four-speaker dialogue data, and enable speaker-turn encoding to develop multi-speaker dialogue generation. Throughout this stage, reference audio is dropped with a probability of 50\%, exposing the model to both reference-conditioned and reference-free generation.

\textbf{2) Dense caption adaptation on clean speech.}
We next train a dense SwanTale model on clean speech data with rich and reliable attribute annotations. At this stage, the Unified MoE layers are replaced with standard dense feed-forward layers. The goal is to teach the model to interpret simple instructions, such as gender, age, and emotion, before introducing sparse routing. Clean bilingual speech with consistent speaker- and utterance-level attributes allows the model to transition from transcript conditioning to caption conditioning while preserving intelligibility, speaker identity, and alignment quality. We additionally incorporate the targeted synthetic subsets described in Section~\ref{sec:data}, improving coverage of elderly speakers, short utterances, and pronunciation-challenging conditions before training on the full caption mixture. In this stage, each sample is trained as an instruct task with a probability of 70\% and as a zero-shot task with a probability of 30\%.

\textbf{3) Full caption-mixture training.}
After the dense model has learned stable caption-conditioned speech generation, we expand the training data to the full SwanData-Caption corpus described in Section~\ref{sec:data} and introduce Unified MoE in the flow-based Transformer. This stage exposes the model to speech, environmental audio, local audio effects, and a broader range of speaker personas. The dense parameters initialize the shared transformation path, while the routed experts learn specialized transformations for speaker variation, environments, and global and local audio effects across the full data mixture.

\textbf{4) High-expressiveness, high-quality SFT.}
The full caption mixture provides broad coverage, but large-scale mixed datasets also contain many samples of moderate quality or limited expressiveness. We therefore continue supervised fine-tuning on the high-expressiveness, high-quality subset selected during data refinement. Samples in this subset must satisfy both automatic waveform-quality thresholds and the human preference audit based on group-wise best--worst comparison described in Section~\ref{sec:data}. The resulting narrower data distribution further improves expressive generation before reward-guided post-training.


\subsection{Reward-guided GRPO Post-training}

The supervised stages give SwanTale broad instruct and zero-shot ability, but recurring errors remain in pronunciation accuracy, generation stability, and caption-conditioned speaker-attribute control. We therefore apply reward-guided post-training to these three targets. The post-training set comprises difficult single-speaker speech cases encountered in advertising, film, television, and animation production, and covers both instruct and zero-shot conditions. Multi-speaker and audio-containing samples are retained through supervised anchor replay rather than assigned unreliable rewards during targeted post-training.


\noindent\textbf{Task-specific reward design.}
We adopt group relative policy optimization (GRPO) \citep{shao2024deepseekmath}. Reward difficulty varies substantially with the target text and condition, so absolute scores from different prompts are not directly comparable. GRPO instead samples multiple candidates under the same condition and optimizes their relative quality without an additional value model.

Both tasks share five speech-side rewards. The phoneme accuracy reward (\texttt{phone\_core}) penalizes substitutions, deletions, and insertions in the recognized phone-tone sequence, while the phoneme-length consistency reward (\texttt{phone\_len}) focuses specifically on deletion and insertion errors:
\begin{equation}
r_{\mathrm{phone}}
=
\exp\left[-\alpha
\frac{w_sS+w_dD+w_iI}{N}\right],
\qquad
r_{\mathrm{len}}
=
\exp\left[-\alpha_{\mathrm{len}}
\frac{w_dD+w_iI}{N}\right],
\end{equation}
where \(S\), \(D\), and \(I\) are the numbers of substitutions, deletions, and insertions, and \(N\) is the phone length. The punctuation-aware pause reward (\texttt{pause\_punct}) checks whether short and long punctuation marks are followed by pauses in their expected duration ranges. The audio-boundary energy reward (\mbox{\texttt{edge\_rms}}) penalizes excessive RMS energy in the first and last 0.2 seconds relative to the whole waveform. The waveform-quality reward (\texttt{quality}) penalizes clipping, abnormal peaks, excessive boundary energy, and high-frequency artifacts.

The final control reward depends on the task. For the instruct task, SwanVerifier predicts age and gender from the generated speech, and the attribute reward measures agreement with the corresponding attributes in the full caption (Appendix~\ref{apx:swanverifier}). For zero-shot generation, speaker identity is supplied by reference audio rather than demographic text; we therefore replace the attribute reward with speaker similarity,
\begin{equation}
r_{\mathrm{sim}}
=
\frac{1+\cos\left(
f_{\mathrm{spk}}(\hat{x}),
f_{\mathrm{spk}}(x_{\mathrm{ref}})
\right)}{2},
\end{equation}
where \(f_{\mathrm{spk}}\) is a frozen WavLM-based ECAPA-TDNN speaker encoder \citep{chen2022wavlm,desplanques2020ecapa}. All reward terms are calibrated to \([0,1]\) before aggregation. Let \(\tau\in\{\mathrm{inst},\mathrm{zero}\}\) denote the task type and \(M_{\tau,m}\) select the rewards applicable to task \(\tau\). The total reward for candidate \(i\) is then computed as follows:
\begin{equation}
R_i^{(\tau)}
=
g_i
\frac{
\sum_m M_{\tau,m}\lambda_m r_{i,m}
}{
\sum_m M_{\tau,m}\lambda_m
},
\end{equation}
where \(g_i\in(0,1]\) is a multiplicative pronunciation gate that downweights samples with severe pronunciation errors. Thus, both instruct and zero-shot tasks share the same pronunciation and stability rewards while switching only the speaker-control reward under a shared aggregation rule.

\noindent\textbf{Stochastic flow policy.}
The original sampler follows the deterministic ODE \(\mathrm{d}\mathbf{x}_t=\mathbf{v}_{\theta}(\mathbf{x}_t,t,c)\mathrm{d}t\), whose transition distribution is degenerate once the initial noise is fixed. Following Flow-GRPO and its TTS adaptation \citep{wang2026flowttsgrpo}, we construct a marginal-preserving SDE
\begin{equation}
\mathrm{d}\mathbf{x}_t
=
\mathbf{b}_{\theta}(\mathbf{x}_t,t,c)\mathrm{d}t
+
\eta(t)\mathrm{d}\mathbf{W}_t,
\qquad
\mathbf{b}_{\theta}
=
\mathbf{v}_{\theta}
+
\frac{\eta(t)^2}{2}\mathbf{s}_{\theta},
\end{equation}
where \(\mathbf{W}_t\) is a Wiener process. Under our noise-to-data convention \(\mathbf{x}_t=(1-t)\mathbf{x}_0+t\mathbf{x}_1\), the rectified-flow score estimate and diffusion schedule are defined as follows:
\begin{equation}
\mathbf{s}_{\theta}(\mathbf{x}_t,t,c)
=
\frac{t\mathbf{v}_{\theta}(\mathbf{x}_t,t,c)-\mathbf{x}_t}{1-t},
\qquad
\eta(t)=a\sqrt{\frac{1-t}{t}},
\end{equation}
with the two endpoints clamped to the nearest interior integration times for numerical stability. For consecutive times \(t_j<t_{j+1}\), Euler--Maruyama gives the Gaussian transition policy
\begin{equation}
\pi_{\theta}(\mathbf{x}_{j+1}\mid\mathbf{x}_j,c)
=
\mathcal{N}\!\left(
\boldsymbol{\mu}_{\theta,j},
\eta(t_j)^2\Delta t_j\mathbf{I}
\right),
\qquad
\boldsymbol{\mu}_{\theta,j}
=
\mathbf{x}_j+
\mathbf{b}_{\theta}(\mathbf{x}_j,t_j,c)\Delta t_j,
\end{equation}
where \(\Delta t_j=t_{j+1}-t_j\). Noise and transition likelihoods are applied only to the generated latent region \(\mathcal{G}\); reference frames remain fixed for zero-shot generation throughout the rollout.

\noindent\textbf{Group-relative policy optimization.}
For condition \(c\), we use a frozen behavior-policy snapshot to sample \(K=8\) trajectories with independent SDE noise realizations, producing a candidate group under the same conditioning input. Then, we score the final waveform generated by each trajectory and normalize the resulting rewards within the group to obtain a group-relative advantage:
\begin{equation}
A_i
=
\frac{
R_i^{(\tau)}-\mu_R(c)
}{
\sigma_R(c)+\epsilon
}.
\end{equation}
During rollout, we store every generated-region transition and its behavior-policy log-probability \(\ell_{\mathrm{old},i,j}\). The current policy recomputes \(\ell_{\theta,i,j}=\log\pi_{\theta}(\mathbf{x}_{i,j+1}\mid\mathbf{x}_{i,j},c)\) using the same SDE kernel. Gaussian log-probabilities are averaged over the valid latent elements in \(\mathcal{G}\) rather than summed. A summed log-likelihood scales with the number of generated elements, so a per-element discrepancy of \(10^{-3}\) already moves \(\rho_{i,j}\) by ten orders of magnitude and saturates the clipping range; the per-element mean keeps \(\rho_{i,j}\) near one and makes a single \(\varepsilon\) comparable across utterance lengths. With \(\rho_{i,j}=\exp(\ell_{\theta,i,j}-\ell_{\mathrm{old},i,j})\), the clipped objective is
\begin{equation}
\mathcal{L}_{\mathrm{GRPO}}
=
-\mathbb{E}_{i,j}
\left[
\min\left(
\rho_{i,j}\tilde{A}_i,\,
\operatorname{clip}(\rho_{i,j},1-\varepsilon,1+\varepsilon)\tilde{A}_i
\right)
\right].
\end{equation}
We directly use \(\tilde{A}_i=\operatorname{clip}(A_i,-A_{\max},A_{\max})\) with \(A_{\max}=2\), and share it across all transitions in trajectory \(i\). \(\rho_{i,j}\) compares the two policies on the same recorded transition, rather than on a denoising surrogate built from a re-noised sample during the clipped policy update.

\noindent\textbf{Capability preservation.}
Targeted single-speaker post-training should not erase the multi-speaker and audio capabilities learned during full-mixture SFT. We use two safeguards. First, a frozen SFT reference policy constrains every GRPO transition. Since the current and reference policies share the same transition variance, their KL divergence has the following closed form:
\begin{equation}
\mathcal{D}_{\mathrm{KL}}
\!\left(
\pi_{\theta}\,\|\,\pi_{\mathrm{ref}}
\right)
=
\frac{
\|\boldsymbol{\mu}_{\theta,j}-\boldsymbol{\mu}_{\mathrm{ref},j}\|_{\mathcal{G}}^2
}{
2\eta(t_j)^2\Delta t_j
}.
\end{equation}
Here, \(\|\cdot\|_{\mathcal{G}}^2\) averages over the valid latent elements in \(\mathcal{G}\), matching the normalization of the transition log-probabilities so that \(\beta_{\mathrm{ref}}\) transfers across utterance lengths. Second, each GRPO update block is followed by supervised anchor steps drawn from the original multi-speaker and audio-containing SFT mixture. These anchor batches use the standard forward noising path and the original masked flow-matching objective, rather than the SDE rollout. The behavior policy is refreshed only after both update blocks, before the next rollout round. The two blocks minimize the following objectives, respectively:
\begin{equation}
\mathcal{L}_{\mathrm{RL}}
=
\mathcal{L}_{\mathrm{GRPO}}
+
\beta_{\mathrm{ref}}
\mathbb{E}_{i,j}
\left[
\mathcal{D}_{\mathrm{KL}}
\!\left(
\pi_{\theta}\,\|\,\pi_{\mathrm{ref}}
\right)
\right]
,\qquad
\mathcal{L}_{\mathrm{anchor}}
=
\mathcal{L}_{\mathrm{flow}}(\mathcal{D}_{\mathrm{anchor}}).
\end{equation}
The reference term limits policy drift on rewarded samples, while anchor replay preserves capabilities outside the coverage of the reward models, including speaker switching, environments, music, and local audio effects.

\subsection{Inference Procedure}

SwanTale uses a unified inference procedure for instruct and zero-shot generation. For instruct generation, the model takes the full caption, and generates the complete latent sequence. For zero-shot generation, it takes a content caption including the content text and reference audio; the reference frames form the latent prompt, and the model generates the unmasked region. In both cases, the text and caption are encoded with the conditioning modules defined above, and speaker-turn labels are constructed from \texttt{<S\{id\}>} tags. The generation region is initialized with Gaussian noise, while prompt frames in the zero-shot setting remain fixed as reference context. SwanTale then solves the flow ODE over the task-specific generation region and decodes the resulting latent trajectory with SwanVAE. 

Content and speaker-turn conditions determine what is spoken and which speaker is active at each turn, whereas caption, reference, and quality conditions control the remaining acoustic attributes. A single guidance scale would couple these two groups of constraints. We therefore use a two-stage decomposed classifier-free guidance (CFG) rule \citep{ho2022classifierfree}. The denoiser is evaluated under a null condition, a text-and-speaker-turn condition, and a task-specific full condition. The guided velocity is
\begin{equation}
\tilde{\mathbf{v}}_t
=
\mathbf{v}_{\emptyset}
+\omega_{\mathrm{text}}(t)(\mathbf{v}_{\mathrm{text}}-\mathbf{v}_{\emptyset})
+\omega_{\mathrm{all}}(t)(\mathbf{v}_{\mathrm{full}}-\mathbf{v}_{\mathrm{text}}),
\end{equation}
where $\omega_{\mathrm{text}}(t)$ guides content and speaker-turn consistency, and $\omega_{\mathrm{all}}(t)$ adds the task-specific full condition. For instruct generation, the full condition contains the full caption and quality flag. For zero-shot generation, it contains the content caption and reference context throughout the conditioned inference trajectory.

Condition strength and numerical resolution need not be uniform over the flow trajectory. Early steps establish coarse speech structure and text--speaker alignment from noise, whereas later steps refine local acoustics. We therefore apply timestep-dependent guidance annealing by setting $\omega_k(t)=\gamma(t)\bar{\omega}_k$ for $k\in\{\mathrm{text},\mathrm{all}\}$, where $\gamma(t)=a+b(1-t)^p$ and $\bar{\omega}_k$ is the corresponding base guidance weight. This applies stronger conditional guidance early and weaker guidance near the data endpoint. Separately, sway sampling \citep{chen2024f5} warps a uniform integration grid $u$ to $t(u)=1-\cos(\pi u/2)$, allocating more Euler steps to the early part of the trajectory.

\section{Experiments}
\label{sec:exp}

\subsection{Implementation Details}

We train SwanVAE on 32 A100 GPUs using roughly 100,000 hours of internal audio spanning speech, singing voice, general audio, and music. Training uses fixed 3.84-s waveform segments, randomly cropped from longer recordings and repeated when necessary for shorter ones. Waveform mixing is applied to 25\% of the crops with $\alpha\sim\mathcal{U}(0.3,0.7)$. Effective-bandwidth augmentation is applied to 1\% of the crops by downsampling them to one of several standard rates from 8 to 44.1~kHz and resampling them back to 48~kHz. The model converts 48~kHz mono waveforms into 96-dimensional continuous latents at 25~Hz. SwanVAE is frozen before training SwanTale. The globally normalized posterior mean $\boldsymbol{\mu}_{\phi}$ is used as the acoustic training target, while generated latents are de-normalized and passed to the frozen decoder for waveform synthesis. SwanVAE contains 407.0M parameters, including 51.7M in the encoder, 0.3M in the variational bottleneck, and 355.0M in the decoder; training-only discriminators and auxiliary heads are excluded from these counts. All SwanTale experiments use the 200k-step checkpoint of the final 96-dimensional model.

All supervised SwanTale stages are trained on 64 A100 GPUs. The first stage trains a 2B-active-parameter zero-shot base model following SwanVoice-style pretraining on 23M hours of single-speaker data and 1.7M hours of two-speaker data. We then perform dense caption adaptation on 70M clean, attribute-rich speech samples for 20k steps. In this stage, we adopt Qwen3.0-Instruct-8B~\citep{yang2025qwen3} as our caption encoder. Next, we train the Unified MoE full-mixture model on 10M SwanData-Caption samples for 10k steps. During caption-mixture training, reference audio is dropped with probability 70\%, so no-reference instruct generation and reference-conditioned zero-shot generation are both observed during training. The final supervised fine-tuning stage uses a 1M-sample subset with high expressiveness and high quality and runs for 4k steps. We use AdamW~\citep{loshchilov2019decoupled} with $\beta_1=0.9$, $\beta_2=0.999$. The peak learning rate is $7.5\times10^{-5}$, and the minimum is $4.0\times10^{-5}$. GRPO post-training uses 8 GPUs and trains for 10 epochs.

For inference, we use the two-stage decomposed CFG rule in Section~\ref{sec:method}, with guidance weights $[1.5, 3.0]$ for the text/speaker branch and the full-condition branch, respectively. We also use the timestep-dependent guidance annealing in Section~\ref{sec:method}, with $(a,b,p)=(0.6,0.6,1.0)$ in every reported experiment.

\subsection{Evaluation Metrics}

\noindent\textbf{SwanVAE.} We evaluate reconstruction on four audio domains: speech, singing voice, general audio, and music. Each domain contains 1,000 clips, drawn from VCTK~\citep{yamagishi2019vctk} for speech, GTSinger~\citep{zhang2024gtsinger} for singing voice, FSD50K~\citep{fonseca2021fsd50k} for general audio, and the MUSDB18-HQ test set~\citep{stoter2018sisec} for music. This gives 4,000 evaluation clips in total. For GTSinger, we compute each metric separately on the Chinese and English subsets and average the two scores with equal weight. We exclude any clips that overlap with the training data. Speech and singing voice share vocal and phonetic structure, so we use the same metric set for both domains: PESQ~\citep{rix2001pesq}, STOI~\citep{taal2011algorithm}, Mel-Cepstral Distortion (MCD)~\citep{kubichek1993mcd}, and ViSQOL~\citep{chinen2020visqol}. For general audio and music, we report ViSQOL and Log-Spectral Distance (LSD)~\citep{gray1976distance}.

\noindent\textbf{Zero-shot task.} SwanBench-Speech~\citep{pan2026swanbench} provides monologue and dialogue test cases paired with reference audio. Timbre Consistency is computed as the mean pairwise cosine similarity among WavLM-TDCNN speaker embeddings~\citep{chen2022wavlm} extracted from sliding windows; for dialogue, we compute the score separately for each speaker and average the resulting per-speaker scores. Reverb Consistency is defined as the standard deviation of window-level SRMR scores~\citep{falk2010srmr}. Sound Fidelity is measured with the reference-free SQUIM-PESQ metric~\citep{kumar2023torchaudiosquim}. Content Error is the unweighted mean of Chinese CER and English WER, with percentages converted to fractions before averaging. SpeechJudge~\citep{zhang2026speechjudge} evaluates Prosodic Coherence based on pauses, speaking rate, and global prosodic consistency. Both expressiveness metrics are scored by Gemini 3 Pro~\citep{pan2026swanbench,googledeepmind2025gemini3pro}. Expressive Richness assesses emotional resonance, character portrayal, and storytelling. Expressive Hierarchy holistically assesses emotional variation, vocal dynamics, and scene appropriateness.

\noindent\textbf{Instruct task.} InstructTTSEval~\citep{huang2025instructttseval} provides speech prompts that specify speaker attributes and delivery in natural language. Acoustic-Parameter Specification (APS) tests direct control of 12 explicit acoustic attributes. Descriptive-Style Directive (DSD) uses free-form descriptions of speaking style. Role-Play (RP) requires the model to infer an appropriate vocal style from a given role or scenario. For all three tasks, we report the official instruction-following accuracy computed by the automatic judge Gemini 2.5 Pro~\citep{googledeepmind2025gemini25pro}.

We also construct \textit{SwanBench-Scene} to evaluate acoustic quality, which is not covered by InstructTTSEval. It contains 180 instruct-TTS instructions: 60 advertising instructions, 60 comic-drama instructions, and 60 general-scene instructions. Annotators rate four dimensions: Overall Expressiveness, Prosodic Naturalness, Audio Fullness, and Scene Appropriateness. Each score is on a 1--5 scale. Each item is evaluated by five professional annotators. Mean MOS is calculated as the arithmetic mean of the four dimension scores.

For hard instruction-based speech and audio generation, we construct \textit{SwanBench-Caption}, a focused evaluation set of 64 cases spanning ambient and localized audio effects, singing voice, background music, multi-language speech, and multi-speaker dialogue. The automatic judge \texttt{gemini-3.5-flash}~\citep{googledeepmind2026gemini35flash} assigns a score from 1 to 5 along three dimensions. Instruction Accuracy measures whether the generated audio correctly realizes the requested content, speaker turns, and sound events. Acoustic Quality assesses clarity, naturalness, and artifacts. Overall Expressiveness assesses emotional delivery, dynamic variation, and scene-level vividness.

\subsection{Baselines}

\noindent\textbf{SwanVAE.} We compare SwanVAE with neural audio codecs and continuous audio autoencoders. To keep the codec comparison consistent across domains, we evaluate DAC~\citep{kumar2023dac}, the 48-kHz EnCodec model~\citep{defossez2022encodec}, and WavTokenizer Large Unify~\citep{ji2024wavtokenizer} on all four test sets. We then add continuous autoencoders according to their intended data domains. For speech and singing voice, these are VoxCPM2 AudioVAE V2~\citep{zhou2026voxcpm2} and MegaTTS 3 WaveVAE~\citep{jiang2025megatts}. For general audio and music, we include Stable Audio Open 1.0~\citep{evans2025stable} and SAME-L~\citep{parker2026same}, the large SAME autoencoder used in Stable Audio 3~\citep{evans2026stable}. ACE-Step Music-DCAE~\citep{gong2025ace} is included for music.

\noindent\textbf{Zero-shot task.} We compare with open-source speech generation systems with reference audio. Monologue baselines include CosyVoice-2~\citep{du2024cosyvoice}, CosyVoice-3~\citep{du2025cosyvoice}, FishSpeech~\citep{fishaudio2024fishspeech}, F5TTS~\citep{chen2024f5}, GLM-TTS~\citep{cui2025glmtts}, IndexTTS-2~\citep{zhou2025indextts2}, MegaTTS-3~\citep{jiang2025megatts}, SparkTTS~\citep{wang2025spark}, VibeVoice~\citep{peng2025vibevoice}, ZipVoice~\citep{zhu2025zipvoice}, and SwanVoice~\citep{li2026swanvoice}. Dialogue baselines include FireRedTTS-2~\citep{xie2025fireredtts}, MoonCast~\citep{ju2025mooncast}, MOSS-TTSD~\citep{mosi_moss_ttsd_2026}, SoulX-Podcast~\citep{xie2025soulx}, VibeVoice~\citep{peng2025vibevoice}, ZipVoice-Dialog~\citep{zhu2025zipvoice}, and SwanVoice~\citep{li2026swanvoice}.

\noindent\textbf{Instruct task.} We compare with representative open-source instruction-following TTS systems: Parler-TTS-large~\citep{lyth2024natural}, VoxInstruct~\citep{zhou2024voxinstruct}, VoiceSculptor~\citep{hu2026voicesculptor}, MiMo-Audio-7B-Instruct~\citep{zhang2025mimoaudio}, Qwen3-TTS-12Hz-1.7B-VD~\citep{qwen2026qwen3tts}, MOSS-VoiceGenerator~\citep{huang2026mossvoicegenerator}, and VoxCPM2~\citep{zhou2026voxcpm2}. On SwanBench-Scene, we add Seedance 2.0~\citep{teamseedance2026seedance2}, using the same input instructions and extracting the audio component from its outputs for evaluation.

\subsection{SwanVAE Evaluation}

We report the latent configurations and reconstruction quality of SwanVAE and the baselines. Table~\ref{tab:swanvae_latent_config} summarizes the sample rate, native latent frame rate, latent size, and nominal rate of each system. Tables~\ref{tab:swanvae_recon_speech_singing} and~\ref{tab:swanvae_recon_general_music} report reconstruction quality on the four test sets described above.

\begin{table}[!htbp]
  \centering
  \caption{Latent configurations of the evaluated audio autoencoders and codecs. Frame rates refer to native encoder outputs before downstream patching. For continuous representations, nominal rates assume 16 bits per scalar (FP16 or BF16); for discrete representations, they count codebook indices only and exclude entropy coding and side information.}
  \label{tab:swanvae_latent_config}
  \small
  \begin{adjustbox}{max width=\textwidth}
  \begin{tabular}{llcccc}
    \toprule
    Model & Representation & Sample Rate & Frame Rate & Latent per Frame & Nominal Rate \\
    \midrule
    ACE-Step Music-DCAE~\citep{gong2025ace}
      & Continuous & 44.1 kHz & 10.77 Hz & 128 scalars & 22.05 kbps \\
    VoxCPM2 AudioVAE V2~\citep{zhou2026voxcpm2}
      & Continuous & $16 \rightarrow 48$ kHz & 25 Hz & 64 scalars & 25.60 kbps \\
    Stable Audio Open 1.0~\citep{evans2025stable}
      & Continuous & 44.1 kHz & 21.53 Hz & 64 scalars & 22.05 kbps \\
    SAME-L~\citep{parker2026same}
      & Continuous & 44.1 kHz & 10.77 Hz & 256 scalars & 44.10 kbps \\
    MegaTTS 3 WaveVAE~\citep{jiang2025megatts}
      & Continuous & 24 kHz & 25 Hz & 32 scalars & 12.80 kbps \\
    \midrule
    DAC~\citep{kumar2023dac}
      & RVQ & 44.1 kHz & 86.13 Hz & $9 \times 10$-bit indices & 7.75 kbps \\
    EnCodec~\citep{defossez2022encodec}
      & RVQ & 48 kHz & 150 Hz & $16 \times 10$-bit indices & 24.00 kbps \\
    WavTokenizer Large Unify~\citep{ji2024wavtokenizer}
      & VQ & 24 kHz & 40 Hz & $1 \times 12$-bit index & 0.48 kbps \\
    \midrule
    SwanVAE (Ours)
      & Continuous & 48 kHz & 25 Hz & 96 scalars & 38.40 kbps \\
    \bottomrule
  \end{tabular}
  \end{adjustbox}
\end{table}

SwanVAE produces 96-dimensional continuous latents at 25 Hz, with one latent step every 40 ms. Under the 16-bit convention in Table~\ref{tab:swanvae_latent_config}, this corresponds to a nominal rate of 38.40 kbps; the same 25-Hz sequence is used as SwanTale's acoustic representation. The following tables evaluate reconstruction quality at this setting across every reported audio domain under the same protocol.

\begin{table}[!htbp]
  \centering
  \caption{Reconstruction quality on the speech and singing voice test sets. Bold and underlined values indicate the best and second-best results among the compared systems within each domain, respectively.}
  \label{tab:swanvae_recon_speech_singing}
  \small
  \begin{tabularx}{\textwidth}{Xcccc}
    \toprule
    Model & PESQ $\uparrow$ & STOI $\uparrow$ & MCD $\downarrow$ & ViSQOL $\uparrow$ \\
    \midrule
    \multicolumn{5}{c}{\textbf{Speech}} \\
    \cmidrule(lr){1-5}
    DAC~\citep{kumar2023dac}
      & \underline{4.1178} & \textbf{0.9693} & \underline{1.1963} & \underline{4.1585} \\
    EnCodec~\citep{defossez2022encodec}
      & 3.1872 & 0.9297 & 1.5147 & 3.5035 \\
    WavTokenizer Large Unify~\citep{ji2024wavtokenizer}
      & 2.1423 & 0.8428 & 2.9393 & 2.3787 \\
    VoxCPM2 AudioVAE V2~\citep{zhou2026voxcpm2}
      & 3.9987 & \underline{0.9690} & 1.2222 & 4.0340 \\
    MegaTTS 3 WaveVAE~\citep{jiang2025megatts}
      & 3.5968 & 0.9507 & 1.5130 & \textbf{4.2348} \\
    \midrule
    SwanVAE (Ours)
      & \textbf{4.1683} & 0.9680 & \textbf{0.9638} & 4.1248 \\
    \midrule
    \multicolumn{5}{c}{\textbf{Singing Voice}} \\
    \cmidrule(lr){1-5}
    DAC~\citep{kumar2023dac}
      & \underline{3.7872} & 0.8627 & \underline{1.9293} & 3.6681 \\
    EnCodec~\citep{defossez2022encodec}
      & 2.6464 & 0.8166 & 2.2392 & 3.3841 \\
    WavTokenizer Large Unify~\citep{ji2024wavtokenizer}
      & 1.9226 & 0.6613 & 5.1885 & 1.6018 \\
    VoxCPM2 AudioVAE V2~\citep{zhou2026voxcpm2}
      & 3.7088 & \underline{0.8838} & 1.9335 & 3.6734 \\
    MegaTTS 3 WaveVAE~\citep{jiang2025megatts}
      & 3.5727 & 0.8620 & 2.0310 & \textbf{4.0013} \\
    \midrule
    SwanVAE (Ours)
      & \textbf{3.9821} & \textbf{0.9001} & \textbf{1.5661} & \underline{3.7085} \\
    \bottomrule
  \end{tabularx}
\end{table}

On speech, SwanVAE achieves the best PESQ and MCD, while remaining close to the strongest baselines on STOI and ViSQOL. On singing voice, it ranks first on PESQ, STOI, and MCD, and second on ViSQOL. The MCD result is consistent across both domains, suggesting that the 25-Hz representation retains vocal spectral structure well despite the low latent frame rate for both vocal domains.

\begin{table}[!htbp]
  \centering
  \caption{Reconstruction quality on the general audio and music test sets. Bold and underlined values indicate the best and second-best results among the compared systems within each domain, respectively.}
  \label{tab:swanvae_recon_general_music}
  \small

  \begin{minipage}[t]{0.48\textwidth}
    \centering
    \begin{tabularx}{\linewidth}{@{}>{\raggedright\arraybackslash}Xcc@{}}
      \toprule
      \multicolumn{3}{c}{\textbf{General Audio}} \\
      \midrule
      Model & ViSQOL $\uparrow$ & LSD $\downarrow$ \\
      \midrule
      DAC~\citep{kumar2023dac}
        & 4.0198 & 0.9589 \\
      EnCodec~\citep{defossez2022encodec}
        & \underline{4.1140} & 0.9761 \\
      WavTokenizer Large Unify~\citep{ji2024wavtokenizer}
        & 2.8595 & 1.0967 \\
      Stable Audio Open 1.0~\citep{evans2025stable}
        & 4.0355 & \textbf{0.9358} \\
      SAME-L~\citep{parker2026same}
        & 3.7541 & 1.0372 \\
      \midrule
      SwanVAE (Ours)
        & \textbf{4.1269} & \underline{0.9455} \\
      \bottomrule
    \end{tabularx}
  \end{minipage}
  \hfill
  \begin{minipage}[t]{0.48\textwidth}
    \centering
    \begin{tabularx}{\linewidth}{@{}>{\raggedright\arraybackslash}Xcc@{}}
      \toprule
      \multicolumn{3}{c}{\textbf{Music}} \\
      \midrule
      Model & ViSQOL $\uparrow$ & LSD $\downarrow$ \\
      \midrule
      DAC~\citep{kumar2023dac}
        & 4.1534 & 0.9196 \\
      EnCodec~\citep{defossez2022encodec}
        & \textbf{4.2976} & \textbf{0.9000} \\
      WavTokenizer Large Unify~\citep{ji2024wavtokenizer}
        & 2.6968 & 1.0612 \\
      Stable Audio Open 1.0~\citep{evans2025stable}
        & 4.1357 & 0.9236 \\
      SAME-L~\citep{parker2026same}
        & 4.0267 & 0.9210 \\
      ACE-Step Music-DCAE~\citep{gong2025ace}
        & 4.1756 & \underline{0.9019} \\
      \midrule
      SwanVAE (Ours)
        & \underline{4.2623} & 0.9172 \\
      \bottomrule
    \end{tabularx}
  \end{minipage}
\end{table}
\FloatBarrier

On general audio, SwanVAE achieves the highest ViSQOL and the second-lowest LSD, behind Stable Audio Open 1.0. On music, EnCodec leads both metrics, while SwanVAE ranks second on ViSQOL and third on LSD. The same SwanVAE checkpoint is used across all four domains, without domain-specific model selection.

\subsection{Zero-Shot Evaluation}

\begin{table}[!htbp]
  \centering
  \caption{Zero-shot monologue and dialogue TTS results on SwanBench-Speech. All scores are reported to two decimal places except Content Error, which is reported to three decimal places. Bold and underlined values indicate the best and second-best results among the compared systems within each setting, respectively.}
  \label{tab:zeroshot_speech}
  \small
  \begin{adjustbox}{max width=\textwidth}
  \begin{tabular}{lccccccc}
    \toprule
    Model & Timbre $\uparrow$ & Reverb $\downarrow$ & Sound Fidelity $\uparrow$ & Content Error $\downarrow$ & SpeechJudge $\uparrow$ & Richness $\uparrow$ & Hierarchy $\uparrow$ \\
    \midrule
    \multicolumn{8}{c}{\textbf{Monologue}} \\
    \cmidrule(lr){1-8}
    CosyVoice-2~\citep{du2024cosyvoice} & 0.93 & 2.37 & 3.58 & 0.106 & 2.81 & 2.02 & 2.59 \\
    CosyVoice-3~\citep{du2025cosyvoice} & 0.93 & 2.73 & 3.80 & 0.077 & 3.26 & 2.64 & 2.47 \\
    FishSpeech~\citep{fishaudio2024fishspeech} & 0.93 & 2.00 & \textbf{4.09} & \textbf{0.066} & \textbf{3.77} & 2.37 & 2.90 \\
    F5TTS~\citep{chen2024f5} & 0.92 & 2.12 & 2.60 & 0.085 & 2.87 & 2.77 & 2.97 \\
    GLM-TTS~\citep{cui2025glmtts} & \underline{0.94} & \textbf{1.64} & 3.90 & 0.074 & 3.28 & 1.57 & 2.39 \\
    IndexTTS-2~\citep{zhou2025indextts2} & 0.93 & \underline{1.77} & 2.78 & 0.077 & 3.63 & 3.32 & 2.94 \\
    MegaTTS-3~\citep{jiang2025megatts} & 0.93 & 2.07 & 3.52 & \underline{0.072} & 3.22 & 2.40 & 3.01 \\
    SparkTTS~\citep{wang2025spark} & 0.92 & 2.04 & 3.53 & 0.314 & 2.35 & 2.23 & 2.22 \\
    VibeVoice~\citep{peng2025vibevoice} & 0.92 & 2.45 & 3.47 & 0.092 & \underline{3.75} & 3.42 & 3.06 \\
    ZipVoice~\citep{zhu2025zipvoice} & 0.89 & 2.10 & 3.53 & 0.213 & 2.97 & 2.11 & 2.05 \\
    \midrule
    SwanVoice~\citep{li2026swanvoice} & 0.93 & 2.06 & 3.60 & 0.172 & 3.56 & \underline{3.81} & \underline{3.62} \\
    SwanTale (Ours) & \textbf{0.95} & 1.83 & \underline{3.95} & 0.086 & \underline{3.75} & \textbf{3.90} & \textbf{3.70} \\
    \midrule
    \multicolumn{8}{c}{\textbf{Dialogue}} \\
    \cmidrule(lr){1-8}
    FireRedTTS-2~\citep{xie2025fireredtts} & 0.91 & 3.54 & 2.54 & 0.148 & 2.93 & 2.52 & 2.65 \\
    MoonCast~\citep{ju2025mooncast} & 0.90 & 3.29 & 2.60 & 0.284 & 2.93 & 2.42 & 2.54 \\
    MOSS-TTSD~\citep{mosi_moss_ttsd_2026} & 0.89 & 3.52 & 2.83 & 0.227 & 2.57 & 3.04 & 2.86 \\
    SoulX-Podcast~\citep{xie2025soulx} & \underline{0.92} & 3.23 & \textbf{3.98} & \textbf{0.101} & \underline{3.89} & 2.80 & 3.15 \\
    VibeVoice~\citep{peng2025vibevoice} & 0.89 & \textbf{2.09} & 2.75 & 0.204 & 3.00 & 3.09 & 2.83 \\
    ZipVoice-Dialog~\citep{zhu2025zipvoice} & 0.90 & 3.49 & 2.48 & \underline{0.116} & 3.46 & 2.88 & 2.93 \\
    \midrule
    SwanVoice~\citep{li2026swanvoice} & \underline{0.92} & \underline{3.02} & \underline{3.77} & 0.145 & 3.70 & \underline{3.62} & \underline{3.71} \\
    SwanTale (Ours) & \textbf{0.94} & 3.21 & 3.73 & 0.120 & \textbf{3.92} & \textbf{3.66} & \textbf{3.85} \\
    \bottomrule
  \end{tabular}
  \end{adjustbox}
\end{table}

Table~\ref{tab:zeroshot_speech} reports SwanBench-Speech results for monologue and two-speaker dialogue zero-shot TTS. SwanTale ranks first in Timbre Consistency, Expressive Richness, and Expressive Hierarchy in both settings, and first on SpeechJudge in the dialogue setting. This indicates that adding and filtering expressive data benefits the model. However, the comparison also reveals room for improvement in content accuracy and audio quality. In the monologue setting, FishSpeech achieves lower Content Error and higher Sound Fidelity; in the dialogue setting, SoulX-Podcast performs best in Sound Fidelity and Content Error among the compared systems.

Moreover, SwanTale consistently improves over SwanVoice. In the monologue zero-shot setting, SwanTale increases Timbre Consistency to 0.95, reduces Content Error to 0.086, and improves SpeechJudge, Expressive Richness, and Expressive Hierarchy to 3.75, 3.90, and 3.70. In the two-speaker dialogue zero-shot setting, the same five metrics improve to 0.94, 0.120, 3.92, 3.66, and 3.85, respectively. These gains result from the complete training recipe: in addition to adding high-expressiveness data, we use ASR-based pronunciation checks to filter the data used for pretraining and supervised fine-tuning (SFT), while reward-conditioned quality control and GRPO further improve generation quality.

\subsection{Instruct Evaluation}

\begin{table}[!htbp]
  \centering
  \caption{Instruct TTS results on InstructTTSEval. Results for all models other than SwanTale are taken from the VoxCPM2 paper~\citep{zhou2026voxcpm2}. Bold and underlined values indicate the best and second-best results.}
  \label{tab:speech_instruct}
  \small
  \begin{tabularx}{\textwidth}{Xcccccc}
    \toprule
    \multirow{2}{*}{Model} & \multicolumn{3}{c}{\textbf{Chinese (ZH)}} & \multicolumn{3}{c}{\textbf{English (EN)}} \\
    \cmidrule(lr){2-4} \cmidrule(lr){5-7}
    & APS $\uparrow$ & DSD $\uparrow$ & RP $\uparrow$ & APS $\uparrow$ & DSD $\uparrow$ & RP $\uparrow$ \\
    \midrule
    Parler-TTS-large~\citep{lyth2024natural} & -- & -- & -- & 60.0 & 45.9 & 31.2 \\
    VoxInstruct~\citep{zhou2024voxinstruct} & 47.5 & 52.3 & 42.6 & 54.9 & 57.0 & 39.3 \\
    VoiceSculptor~\citep{hu2026voicesculptor} & 75.7 & 64.7 & 61.5 & -- & -- & -- \\
    MiMo-Audio-7B-Instruct~\citep{zhang2025mimoaudio} & 75.7 & 74.3 & 61.5 & 80.6 & 77.6 & 59.5 \\
    Qwen3-TTS-12Hz-1.7B-VD~\citep{qwen2026qwen3tts} & \underline{85.2} & \textbf{81.1} & \underline{65.1} & \underline{82.9} & \underline{82.4} & 68.4 \\
    MOSS-VoiceGenerator~\citep{huang2026mossvoicegenerator} & 78.0 & 80.0 & \textbf{74.0} & 68.2 & 82.0 & \underline{68.7} \\
    VoxCPM2~\citep{zhou2026voxcpm2} & \underline{85.2} & 71.5 & 60.8 & \textbf{84.2} & \textbf{83.2} & \textbf{71.4} \\
    \midrule
    SwanTale (Ours) & \textbf{86.1} & \underline{80.1} & 64.1 & \textbf{84.2} & 79.2 & 63.6 \\
    \bottomrule
  \end{tabularx}
\end{table}

\begin{table}[!htbp]
  \centering
  \caption{Results on SwanBench-Scene. Mean MOS is the arithmetic mean of the other four metrics. Bold and underlined values indicate the best and second-best results.}
  \label{tab:swanbench_scene}
  \small
  \begin{adjustbox}{max width=\textwidth}
  \begin{tabular}{lccccc}
    \toprule
    Model & Mean MOS $\uparrow$ & Overall Exp. $\uparrow$ & Prosodic Nat. $\uparrow$ & Audio Fullness $\uparrow$ & Scene App. $\uparrow$ \\
    \midrule
    \multicolumn{6}{c}{\textbf{Advertising}} \\
    \cmidrule(lr){1-6}
    MOSS-VoiceGenerator~\citep{huang2026mossvoicegenerator} & 3.13 & 2.89 & 3.10 & 3.75 & 2.76 \\
    MiMo-Audio-7B-Instruct~\citep{zhang2025mimoaudio} & 3.13 & 2.96 & 3.17 & 3.49 & 2.89 \\
    Seedance 2.0~\citep{teamseedance2026seedance2} & 3.38 & 3.25 & 3.34 & 3.73 & 3.22 \\
    Qwen3-TTS-12Hz-1.7B-VD~\citep{qwen2026qwen3tts} & \underline{3.71} & \underline{3.62} & \underline{3.73} & \underline{4.12} & \underline{3.38} \\
    SwanTale (Ours) & \textbf{3.88} & \textbf{3.76} & \textbf{3.82} & \textbf{4.21} & \textbf{3.73} \\
    \midrule
    \multicolumn{6}{c}{\textbf{Comic Drama}} \\
    \cmidrule(lr){1-6}
    MiMo-Audio-7B-Instruct & 3.23 & 3.06 & 3.24 & 3.72 & 2.89 \\
    MOSS-VoiceGenerator & 4.06 & 3.94 & 4.11 & 4.33 & 3.87 \\
    Qwen3-TTS-12Hz-1.7B-VD & \underline{4.35} & 4.20 & 4.35 & \textbf{4.63} & 4.21 \\
    Seedance 2.0 & \underline{4.35} & \underline{4.30} & \underline{4.42} & 4.45 & \underline{4.23} \\
    SwanTale (Ours) & \textbf{4.45} & \textbf{4.36} & \textbf{4.47} & \underline{4.60} & \textbf{4.36} \\
    \midrule
    \multicolumn{6}{c}{\textbf{General Scene}} \\
    \cmidrule(lr){1-6}
    MiMo-Audio-7B-Instruct & 3.28 & 3.17 & 3.39 & 3.49 & 3.05 \\
    MOSS-VoiceGenerator & 3.62 & 3.42 & 3.65 & 4.10 & 3.29 \\
    Seedance 2.0 & 4.14 & \underline{4.11} & \underline{4.18} & 4.23 & \underline{4.05} \\
    Qwen3-TTS-12Hz-1.7B-VD & \underline{4.22} & 4.05 & \underline{4.18} & \underline{4.59} & 4.04 \\
    SwanTale (Ours) & \textbf{4.34} & \textbf{4.24} & \textbf{4.30} & \textbf{4.60} & \textbf{4.21} \\
    \midrule
    \multicolumn{6}{c}{\textbf{Overall}} \\
    \cmidrule(lr){1-6}
    MiMo-Audio-7B-Instruct & 3.19 & 3.04 & 3.24 & 3.55 & 2.93 \\
    MOSS-VoiceGenerator & 3.48 & 3.29 & 3.49 & 3.98 & 3.17 \\
    Seedance 2.0 & 3.86 & 3.78 & 3.87 & 4.07 & 3.73 \\
    Qwen3-TTS-12Hz-1.7B-VD & \underline{4.09} & \underline{3.96} & \underline{4.09} & \underline{4.45} & \underline{3.88} \\
    SwanTale (Ours) & \textbf{4.22} & \textbf{4.12} & \textbf{4.20} & \textbf{4.47} & \textbf{4.10} \\
    \bottomrule
  \end{tabular}
  \end{adjustbox}
\end{table}

\noindent\textbf{InstructTTSEval.} As shown in Table \ref{tab:speech_instruct}, SwanTale is strongest on explicit acoustic control and performs strongly on Chinese descriptive-style instructions. It ranks first on Chinese APS (86.1), ties for first on English APS (84.2), and ranks second on Chinese DSD (80.1). APS and DSD closely match the detailed acoustic and speaking-style descriptions used in our caption data. Meanwhile, SwanTale's Chinese and English scores are close within each task, showing balanced performance across languages. English DSD is less competitive in the cross-system comparison: several baselines improve markedly from Chinese to English, while SwanTale stays at a similar level. RP is also a weakness in both languages. It requires the model to map an occupation, character archetype, or scenario to a recognizable vocal performance. Our caption annotation and current style matrices may not cover enough long-tail role-imitation styles, limiting generalization to unseen roles and scenarios. Therefore, improving RP will require broader caption and style-matrix coverage of roles, occupations, character archetypes, and situations in practical instruction settings.

\noindent\textbf{SwanBench-Scene.} Table~\ref{tab:swanbench_scene} summarizes the SwanBench-Scene results. SwanTale achieves the highest overall Mean MOS (4.22) and the highest overall scores for Overall Expressiveness (4.12), Prosodic Naturalness (4.20), Audio Fullness (4.47), and Scene Appropriateness (4.10). It also obtains the highest Mean MOS for advertising (3.88), comic drama (4.45), and general scenes (4.34). Its only second-place dimension score is Audio Fullness in comic drama (4.60) among all evaluated systems.

\begin{table}[!htbp]
  \centering
  \caption{Results on SwanBench-Caption. All metrics are scored on a 1--5 scale by \texttt{gemini-3.5-flash}; higher is better. \textit{32B CE} replaces the default Qwen3.0-Instruct-8B caption encoder with Qwen3.0-Instruct-32B~\citep{yang2025qwen3}. }
  \label{tab:caption_instruct}
  \small
  \begin{tabularx}{\textwidth}{Xccc}
    \toprule
    Setting & Instruction Accuracy $\uparrow$ & Acoustic Quality $\uparrow$ & Overall Expressiveness $\uparrow$ \\
    \midrule
    SwanTale w/o MoE & 3.02 & 4.09 & 3.56 \\
    SwanTale & 3.39 & 4.31 & 3.82 \\
    SwanTale w/ 32B CE & 3.70 & 4.34 & 3.98 \\
    \bottomrule
  \end{tabularx}
\end{table}

\noindent\textbf{SwanBench-Caption.} Table~\ref{tab:caption_instruct} reports the ablation results on SwanBench-Caption. Removing Unified MoE lowers Instruction Accuracy from 3.39 to 3.02, Acoustic Quality from 4.31 to 4.09, and Overall Expressiveness from 3.82 to 3.56. The declines across all three metrics show that Unified MoE benefits instruction realization, acoustic quality, and expressiveness. Scaling the caption encoder from 8B to 32B further raises the three scores to 3.70, 4.34, and 3.98, respectively, with the largest increase in Instruction Accuracy.

Overall, the three instruct evaluations reveal complementary strengths. SwanTale leads APS and performs strongly on Chinese DSD, while English DSD and RP are less competitive. SwanBench-Scene demonstrates high perceptual quality across diverse scenarios, and the SwanBench-Caption ablations show gains from Unified MoE and increased caption-encoder capacity on complex speech-and-audio instructions.

\section{Conclusion}
\label{sec:con}

This work presents SwanTale, a unified model for multi-speaker speech and audio generation across instruct and zero-shot tasks. SwanData-Caption provides the required supervision through targeted data coverage, speech-aware preprocessing, multi-level caption annotation, and quality filtering. SwanTale combines SwanVAE, a flow-based Transformer with reward-conditioned quality control, Engram conditioning, and Unified MoE, together with curriculum learning and GRPO post-training, to progressively adapt a unified generator to both caption and reference-audio conditioning. The resulting model can design speaker voices from natural-language descriptions, reuse these voices through reference audio, and jointly generate speech, environments, and local audio effects in a single waveform through one unified generation process.

Experiments show that SwanTale leads on multiple key zero-shot and instruction-following metrics. It achieves the best expressiveness scores in both zero-shot and instruct tasks and supports complex instruct generation involving multi-speaker speech and audio within the same model.

Several challenges remain for both instruct and zero-shot generation. First, complex background music generation remains difficult, particularly when the music must change in type or transition in response to different emotions. Second, long-form instruct generation is still challenging, especially for complex multi-speaker scenes longer than two minutes that also contain audio effects. Third, precise local style control remains difficult. This includes continuous emotional changes for a specified speaker, accurate control over emphasis and speaking rhythm, and well-timed pauses and audio effects, all of which pose challenges for both data annotation and model design. Beyond generation, editing the spoken content, speaker identity, and emotion of existing audio is another important problem~\citep{pan2026audioediting}. A model that unifies audio generation and editing is therefore a clear direction for future research within a common framework.

\clearpage
\bibliographystyle{plainnat}
\bibliography{custom}

\begin{thebibliography}{119}
\providecommand{\natexlab}[1]{#1}
\providecommand{\url}[1]{\texttt{#1}}
\expandafter\ifx\csname urlstyle\endcsname\relax
  \providecommand{\doi}[1]{doi: #1}\else
  \providecommand{\doi}{doi: \begingroup \urlstyle{rm}\Url}\fi

\bibitem[An et~al.(2024)An, Chen, Deng, Du, Gao, Gao, Gu, He, Hu, Hu, et~al.]{an2024funaudiollm}
Keyu An, Qian Chen, Chong Deng, Zhihao Du, Changfeng Gao, Zhifu Gao, Yue Gu, Ting He, Hangrui Hu, Kai Hu, et~al.
\newblock Funaudiollm: Voice understanding and generation foundation models for natural interaction between humans and llms.
\newblock \emph{arXiv preprint arXiv:2407.04051}, 2024.

\bibitem[Anjok07 and aufr33(2020)]{ultimatevocalremovergui}
Anjok07 and aufr33.
\newblock Ultimate vocal remover.
\newblock \emph{GitHub repository}, 2020.
\newblock URL \url{https://github.com/Anjok07/ultimatevocalremovergui}.

\bibitem[Bai et~al.(2023)Bai, Bai, Chu, Cui, Dang, Deng, Fan, Ge, Han, Huang, et~al.]{bai2023qwen}
Jinze Bai, Shuai Bai, Yunfei Chu, Zeyu Cui, Kai Dang, Xiaodong Deng, Yang Fan, Wenbin Ge, Yu~Han, Fei Huang, et~al.
\newblock Qwen technical report.
\newblock \emph{arXiv preprint arXiv:2309.16609}, 2023.

\bibitem[Bain et~al.(2023)Bain, Huh, Han, and Zisserman]{bain2023whisperx}
Max Bain, Jaesung Huh, Tengda Han, and Andrew Zisserman.
\newblock Whisperx: Time-accurate speech transcription of long-form audio.
\newblock \emph{arXiv preprint arXiv:2303.00747}, 2023.

\bibitem[Bengio et~al.(2009)Bengio, Louradour, Collobert, and Weston]{bengio2009curriculum}
Yoshua Bengio, J{\'e}r{\^o}me Louradour, Ronan Collobert, and Jason Weston.
\newblock Curriculum learning.
\newblock In \emph{Proceedings of the 26th Annual International Conference on Machine Learning}, pages 41--48. ACM, 2009.

\bibitem[{ByteDance Seed Team}(2026)]{bytedanceseed2026seed2}
{ByteDance Seed Team}.
\newblock {Seed2.0}, 2026.
\newblock URL \url{https://seed.bytedance.com/en/seed2}.
\newblock Model website.

\bibitem[{BytePlus}(2026)]{byteplus2026seedasr2}
{BytePlus}.
\newblock {Seed Speech ASR 2.0 Documentation}.
\newblock \emph{BytePlus documentation}, 2026.
\newblock URL \url{https://docs.byteplus.com/en/docs/byteplusvoice/speechtotextv2}.
\newblock Accessed: 2026-06-01.

\bibitem[Chen et~al.(2026)Chen, Zhang, Wang, Dai, Ma, and Wu]{chen2026flexivoice}
Dekun Chen, Xueyao Zhang, Yuancheng Wang, Kenan Dai, Li~Ma, and Zhizheng Wu.
\newblock {FlexiVoice}: Enabling flexible style control in zero-shot {TTS} with natural language instructions.
\newblock In \emph{International Conference on Learning Representations}, 2026.

\bibitem[Chen et~al.(2021)Chen, Chai, Wang, Du, Zhang, Weng, Su, Povey, Trmal, Zhang, et~al.]{chen2021gigaspeech}
Guoguo Chen, Shuzhou Chai, Guanbo Wang, Jiayu Du, Wei-Qiang Zhang, Chao Weng, Dan Su, Daniel Povey, Jan Trmal, Junbo Zhang, et~al.
\newblock Gigaspeech: An evolving, multi-domain asr corpus with 10,000 hours of transcribed audio.
\newblock \emph{arXiv preprint arXiv:2106.06909}, 2021.

\bibitem[Chen et~al.(2022)Chen, Wang, Chen, Wu, Liu, Chen, Li, Kanda, Yoshioka, Xiao, et~al.]{chen2022wavlm}
Sanyuan Chen, Chengyi Wang, Zhengyang Chen, Yu~Wu, Shujie Liu, Zhuo Chen, Jinyu Li, Naoyuki Kanda, Takuya Yoshioka, Xiong Xiao, et~al.
\newblock Wavlm: Large-scale self-supervised pre-training for full stack speech processing.
\newblock \emph{IEEE Journal of Selected Topics in Signal Processing}, 16\penalty0 (6):\penalty0 1505--1518, 2022.

\bibitem[Chen et~al.(2025{\natexlab{a}})Chen, Zheng, Wang, Cheng, Zhu, Huang, Deng, Chen, Zhang, Wang, et~al.]{chen20253d}
Yafeng Chen, Siqi Zheng, Hui Wang, Luyao Cheng, Tinglong Zhu, Rongjie Huang, Chong Deng, Qian Chen, Shiliang Zhang, Wen Wang, et~al.
\newblock 3d-speaker-toolkit: An open-source toolkit for multimodal speaker verification and diarization.
\newblock In \emph{ICASSP 2025-2025 IEEE International Conference on Acoustics, Speech and Signal Processing (ICASSP)}, pages 1--5. IEEE, 2025{\natexlab{a}}.

\bibitem[Chen et~al.(2025{\natexlab{b}})Chen, Wang, Wang, Chen, He, Zhou, Yang, Wang, Lin, and Qin]{chen2025seniortalk}
Yang Chen, Hui Wang, Shiyao Wang, Junyang Chen, Jiabei He, Jiaming Zhou, Xi~Yang, Yequan Wang, Yonghua Lin, and Yong Qin.
\newblock {SeniorTalk}: A chinese conversation dataset with rich annotations for super-aged seniors.
\newblock \emph{arXiv preprint arXiv:2503.16578}, 2025{\natexlab{b}}.
\newblock URL \url{https://arxiv.org/abs/2503.16578}.

\bibitem[Chen et~al.(2024)Chen, Niu, Ma, Deng, Wang, Zhao, Yu, and Chen]{chen2024f5}
Yushen Chen, Zhikang Niu, Ziyang Ma, Keqi Deng, Chunhui Wang, Jian Zhao, Kai Yu, and Xie Chen.
\newblock F5-tts: A fairytaler that fakes fluent and faithful speech with flow matching.
\newblock \emph{arXiv preprint arXiv:2410.06885}, 2024.

\bibitem[Cheng et~al.(2026)Cheng, Zeng, Dai, Chen, Wang, Xie, Huang, Yu, Hao, Li, Zhang, Zhang, Zhao, and Liang]{cheng2026conditionalmemory}
Xin Cheng, Wangding Zeng, Damai Dai, Qinyu Chen, Bingxuan Wang, Zhenda Xie, Kezhao Huang, Xingkai Yu, Zhewen Hao, Yukun Li, Han Zhang, Huishuai Zhang, Dongyan Zhao, and Wenfeng Liang.
\newblock Conditional memory via scalable lookup: A new axis of sparsity for large language models.
\newblock \emph{arXiv preprint arXiv:2601.07372}, 2026.

\bibitem[Chinen et~al.(2020)Chinen, Lim, Skoglund, Gureev, O'Gorman, and Hines]{chinen2020visqol}
Michael Chinen, Felicia S.~C. Lim, Jan Skoglund, Nikita Gureev, Feargus O'Gorman, and Andrew Hines.
\newblock {ViSQOL v3}: An open source production ready objective speech and audio metric.
\newblock In \emph{2020 Twelfth International Conference on Quality of Multimedia Experience (QoMEX)}, pages 1--6, 2020.
\newblock \doi{10.1109/QoMEX48832.2020.9123150}.

\bibitem[Chung et~al.(2019)Chung, Hsu, Tang, and Glass]{chung2019apc}
Yu-An Chung, Wei-Ning Hsu, Hao Tang, and James Glass.
\newblock An unsupervised autoregressive model for speech representation learning.
\newblock \emph{arXiv preprint arXiv:1904.03240}, 2019.

\bibitem[Cui et~al.(2025)Cui, Yang, Li, Tian, Ma, Zhang, Chen, Yang, Cheng, Zhou, Yu, Gu, and Tang]{cui2025glmtts}
Jiayan Cui, Zhihan Yang, Naihan Li, Jiankun Tian, Xingyu Ma, Yi~Zhang, Guangyu Chen, Runxuan Yang, Yuqing Cheng, Yizhi Zhou, Guochen Yu, Xiaotao Gu, and Jie Tang.
\newblock {GLM-4-Voice}: Towards intelligent and human-like end-to-end spoken chatbot.
\newblock \emph{arXiv preprint arXiv:2507.01006}, 2025.

\bibitem[Dai et~al.(2024)Dai, Deng, Zhao, Xu, Gao, Chen, Li, Zeng, Yu, Wu, Xie, Li, Huang, Luo, Ruan, Sui, and Liang]{dai2024deepseekmoe}
Damai Dai, Chengqi Deng, Chenggang Zhao, R.~X. Xu, Huazuo Gao, Deli Chen, Jiashi Li, Wangding Zeng, Xingkai Yu, Y.~Wu, Zhenda Xie, Y.~K. Li, Panpan Huang, Fuli Luo, Chong Ruan, Zhifang Sui, and Wenfeng Liang.
\newblock {DeepSeekMoE}: Towards ultimate expert specialization in mixture-of-experts language models.
\newblock \emph{arXiv preprint arXiv:2401.06066}, 2024.

\bibitem[Defossez et~al.(2022)Defossez, Copet, Synnaeve, and Adi]{defossez2022encodec}
Alexandre Defossez, Jade Copet, Gabriel Synnaeve, and Yossi Adi.
\newblock High fidelity neural audio compression.
\newblock \emph{arXiv preprint arXiv:2210.13438}, 2022.

\bibitem[Desplanques et~al.(2020)Desplanques, Thienpondt, and Demuynck]{desplanques2020ecapa}
Brecht Desplanques, Jenthe Thienpondt, and Kris Demuynck.
\newblock {ECAPA-TDNN}: Emphasized channel attention, propagation and aggregation in {TDNN} based speaker verification.
\newblock In \emph{Interspeech}, pages 3830--3834, 2020.

\bibitem[Du et~al.(2024)Du, Chen, Zhang, Hu, Lu, Yang, Hu, Zheng, Gu, Ma, et~al.]{du2024cosyvoice}
Zhihao Du, Qian Chen, Shiliang Zhang, Kai Hu, Heng Lu, Yexin Yang, Hangrui Hu, Siqi Zheng, Yue Gu, Ziyang Ma, et~al.
\newblock Cosyvoice: A scalable multilingual zero-shot text-to-speech synthesizer based on supervised semantic tokens.
\newblock \emph{arXiv preprint arXiv:2407.05407}, 2024.

\bibitem[Du et~al.(2025)Du, Gao, Wang, Yu, Zhao, Wang, Lv, Wang, Ni, Shi, et~al.]{du2025cosyvoice}
Zhihao Du, Changfeng Gao, Yuxuan Wang, Fan Yu, Tianyu Zhao, Hao Wang, Xiang Lv, Hui Wang, Chongjia Ni, Xian Shi, et~al.
\newblock Cosyvoice 3: Towards in-the-wild speech generation via scaling-up and post-training.
\newblock \emph{arXiv preprint arXiv:2505.17589}, 2025.

\bibitem[Evans et~al.(2025)Evans, Parker, Carr, Zukowski, Taylor, and Pons]{evans2025stable}
Zach Evans, Julian~D Parker, CJ~Carr, Zack Zukowski, Josiah Taylor, and Jordi Pons.
\newblock Stable audio open.
\newblock In \emph{ICASSP 2025-2025 IEEE International Conference on Acoustics, Speech and Signal Processing (ICASSP)}, pages 1--5. IEEE, 2025.

\bibitem[Evans et~al.(2026)Evans, Parker, Rice, Carr, Zukowski, Taylor, and Pons]{evans2026stable}
Zach Evans, Julian~D Parker, Matthew Rice, CJ~Carr, Zack Zukowski, Josiah Taylor, and Jordi Pons.
\newblock Stable audio 3.
\newblock \emph{arXiv preprint arXiv:2605.17991}, 2026.

\bibitem[Falk et~al.(2010)Falk, Zheng, and Chan]{falk2010srmr}
Tiago~H. Falk, Chenxi Zheng, and Wai-Yip Chan.
\newblock A non-intrusive quality and intelligibility measure of reverberant and dereverberated speech.
\newblock \emph{IEEE Transactions on Audio, Speech, and Language Processing}, 18\penalty0 (7):\penalty0 1766--1774, 2010.
\newblock \doi{10.1109/TASL.2010.2052247}.

\bibitem[Fedus et~al.(2022)Fedus, Zoph, and Shazeer]{fedus2022switch}
William Fedus, Barret Zoph, and Noam Shazeer.
\newblock Switch transformers: Scaling to trillion parameter models with simple and efficient sparsity.
\newblock \emph{Journal of Machine Learning Research}, 23\penalty0 (120):\penalty0 1--39, 2022.

\bibitem[{Fish Audio Team}(2024)]{fishaudio2024fishspeech}
{Fish Audio Team}.
\newblock {Fish-Speech}: Leveraging large language models for advanced multilingual text-to-speech synthesis.
\newblock \emph{arXiv preprint arXiv:2411.01156}, 2024.

\bibitem[Fonseca et~al.(2021)Fonseca, Favory, Pons, Font, and Serra]{fonseca2021fsd50k}
Eduardo Fonseca, Xavier Favory, Jordi Pons, Frederic Font, and Xavier Serra.
\newblock Fsd50k: an open dataset of human-labeled sound events.
\newblock \emph{IEEE/ACM Transactions on Audio, Speech, and Language Processing}, 30:\penalty0 829--852, 2021.

\bibitem[Gong et~al.(2025)Gong, Zhao, Wang, Xu, and Guo]{gong2025ace}
Junmin Gong, Sean Zhao, Sen Wang, Shengyuan Xu, and Joe Guo.
\newblock Ace-step: A step towards music generation foundation model.
\newblock \emph{arXiv preprint arXiv:2506.00045}, 2025.

\bibitem[{Google DeepMind}(2025{\natexlab{a}})]{googledeepmind2025gemini25pro}
{Google DeepMind}.
\newblock {Gemini 2.5 Pro} model card, 2025{\natexlab{a}}.
\newblock URL \url{https://storage.googleapis.com/deepmind-media/Model-Cards/Gemini-2-5-Pro-Model-Card.pdf}.
\newblock Model card.

\bibitem[{Google DeepMind}(2025{\natexlab{b}})]{googledeepmind2025gemini3pro}
{Google DeepMind}.
\newblock {Gemini 3 Pro} model card, 2025{\natexlab{b}}.
\newblock URL \url{https://deepmind.google/models/model-cards/gemini-3-pro/}.
\newblock Model card.

\bibitem[{Google DeepMind}(2026)]{googledeepmind2026gemini35flash}
{Google DeepMind}.
\newblock {Gemini 3.5 Flash} model card, 2026.
\newblock URL \url{https://deepmind.google/models/model-cards/gemini-3-5-flash/}.
\newblock Model card.

\bibitem[Gray and Markel(1976)]{gray1976distance}
Augustine~H. Gray and John~D. Markel.
\newblock Distance measures for speech processing.
\newblock \emph{IEEE Transactions on Acoustics, Speech, and Signal Processing}, 24\penalty0 (5):\penalty0 380--391, 1976.
\newblock \doi{10.1109/TASSP.1976.1162849}.

\bibitem[Guo et~al.(2025{\natexlab{a}})Guo, Pan, Zhu, Hu, Zhang, Tang, Yang, Wang, Zhang, Wang, Chen, Xu, Xu, Fan, Chen, Yu, Huang, Wu, and Zhao]{guo2025mrsaudio}
Wenxiang Guo, Changhao Pan, Zhiyuan Zhu, Xintong Hu, Yu~Zhang, Li~Tang, Rui Yang, Han Wang, Zongbao Zhang, Yuhan Wang, Yixuan Chen, Hankun Xu, Ke~Xu, Pengfei Fan, Zhetao Chen, Yanhao Yu, Qiange Huang, Fei Wu, and Zhou Zhao.
\newblock {MRSAudio}: A large-scale multimodal recorded spatial audio dataset with refined annotations.
\newblock In \emph{Advances in Neural Information Processing Systems}, 2025{\natexlab{a}}.

\bibitem[Guo et~al.(2025{\natexlab{b}})Guo, Zhang, Pan, Huang, Tang, Li, Hong, Wang, and Zhao]{guo2025techsinger}
Wenxiang Guo, Yu~Zhang, Changhao Pan, Rongjie Huang, Li~Tang, Ruiqi Li, Zhiqing Hong, Yongqi Wang, and Zhou Zhao.
\newblock Techsinger: Technique controllable multilingual singing voice synthesis via flow matching.
\newblock \emph{arXiv preprint arXiv:2502.12572}, 2025{\natexlab{b}}.

\bibitem[Guo et~al.(2025{\natexlab{c}})Guo, Zhang, Pan, Zhu, Li, Chen, Xu, Wu, and Zhao]{guo2025stars}
Wenxiang Guo, Yu~Zhang, Changhao Pan, Zhiyuan Zhu, Ruiqi Li, ZheTao Chen, Wenhao Xu, Fei Wu, and Zhou Zhao.
\newblock {STARS}: A unified framework for singing transcription, alignment, and refined style annotation.
\newblock In \emph{Findings of the Association for Computational Linguistics: ACL 2025}, pages 15081--15093, Vienna, Austria, 2025{\natexlab{c}}. Association for Computational Linguistics.

\bibitem[Ho and Salimans(2022)]{ho2022classifierfree}
Jonathan Ho and Tim Salimans.
\newblock Classifier-free diffusion guidance.
\newblock \emph{arXiv preprint arXiv:2207.12598}, 2022.

\bibitem[Hu et~al.(2026)Hu, Chen, Ma, Guo, Zhan, Li, Zhang, Xia, Zhang, Tian, Wang, Liang, Guo, Yang, Wu, Zhang, Zhu, Xie, Xie, Zhang, Liu, and Xie]{hu2026voicesculptor}
Jingbin Hu, Huakang Chen, Linhan Ma, Dake Guo, Qirui Zhan, Wenhao Li, Haoyu Zhang, Kangxiang Xia, Ziyu Zhang, Wenjie Tian, Chengyou Wang, Jinrui Liang, Shuhan Guo, Zihang Yang, Bengu Wu, Binbin Zhang, Pengcheng Zhu, Pengyuan Xie, Chuan Xie, Qiang Zhang, Jie Liu, and Lei Xie.
\newblock {VoiceSculptor}: Your voice, designed by you.
\newblock \emph{arXiv preprint arXiv:2601.10629}, 2026.

\bibitem[Hu et~al.(2025)Hu, Puvvada, Rastorgueva, Chen, Huang, Ding, Dhawan, Xu, Balam, and Ginsburg]{hu2025word}
Ke~Hu, Krishna Puvvada, Elena Rastorgueva, Zhehuai Chen, He~Huang, Shuoyang Ding, Kunal Dhawan, Hainan Xu, Jagadeesh Balam, and Boris Ginsburg.
\newblock Word level timestamp generation for automatic speech recognition and translation.
\newblock \emph{arXiv preprint arXiv:2505.15646}, 2025.

\bibitem[Huang et~al.(2023{\natexlab{a}})Huang, Kong, et~al.]{mozillazg2023pypinyin}
Huang Huang, Xiaoquan Kong, et~al.
\newblock python-pinyin: pypinyin, 2023{\natexlab{a}}.
\newblock URL \url{https://github.com/mozillazg/python-pinyin}.
\newblock Software, version 0.48.0.

\bibitem[Huang et~al.(2025)Huang, Tu, Fan, Yang, Zhang, Li, Fei, Cheng, and Qiu]{huang2025instructttseval}
Kexin Huang, Qian Tu, Liwei Fan, Chenchen Yang, Dong Zhang, Shimin Li, Zhaoye Fei, Qinyuan Cheng, and Xipeng Qiu.
\newblock {InstructTTSEval}: Benchmarking complex natural-language instruction following in text-to-speech systems.
\newblock \emph{arXiv preprint arXiv:2506.16381}, 2025.

\bibitem[Huang et~al.(2026)Huang, Fan, Jiang, Jiang, Tu, Zhu, Zhang, Zhao, Yang, Fei, Li, Yang, Cheng, and Qiu]{huang2026mossvoicegenerator}
Kexin Huang, Liwei Fan, Botian Jiang, Yaozhou Jiang, Qian Tu, Jie Zhu, Yuqian Zhang, Yiwei Zhao, Chenchen Yang, Zhaoye Fei, Shimin Li, Xiaogui Yang, Qinyuan Cheng, and Xipeng Qiu.
\newblock {MOSS-VoiceGenerator}: Create realistic voices with natural language descriptions.
\newblock \emph{arXiv preprint arXiv:2603.28086}, 2026.

\bibitem[Huang et~al.(2022)Huang, Ren, Liu, Cui, and Zhao]{huang2022generspeech}
Rongjie Huang, Yi~Ren, Jinglin Liu, Chenye Cui, and Zhou Zhao.
\newblock Generspeech: Towards style transfer for generalizable out-of-domain text-to-speech.
\newblock \emph{Advances in Neural Information Processing Systems (NeurIPS)}, 2022.

\bibitem[Huang et~al.(2023{\natexlab{b}})Huang, Huang, Yang, Ren, Liu, Li, Ye, Liu, Yin, and Zhao]{huang2023make}
Rongjie Huang, Jiawei Huang, Dongchao Yang, Yi~Ren, Luping Liu, Mingze Li, Zhenhui Ye, Jinglin Liu, Xiang Yin, and Zhou Zhao.
\newblock Make-an-audio: Text-to-audio generation with prompt-enhanced diffusion models.
\newblock In \emph{International Conference on Machine Learning}, pages 13916--13932. PMLR, 2023{\natexlab{b}}.

\bibitem[Jang et~al.(2016)Jang, Gu, and Poole]{jang2016categorical}
Eric Jang, Shixiang Gu, and Ben Poole.
\newblock Categorical reparameterization with gumbel-softmax.
\newblock \emph{arXiv preprint arXiv:1611.01144}, 2016.

\bibitem[Jang et~al.(2021)Jang, Lim, Yoon, Kim, and Kim]{jang2021univnet}
Won Jang, Dan Lim, Jaesam Yoon, Bongwan Kim, and Juntae Kim.
\newblock Univnet: A neural vocoder with multi-resolution spectrogram discriminators for high-fidelity waveform generation.
\newblock \emph{arXiv preprint arXiv:2106.07889}, 2021.

\bibitem[Ji et~al.(2024)Ji, Jiang, Wang, Chen, Fang, Zuo, Yang, Cheng, Wang, Li, et~al.]{ji2024wavtokenizer}
Shengpeng Ji, Ziyue Jiang, Wen Wang, Yifu Chen, Minghui Fang, Jialong Zuo, Qian Yang, Xize Cheng, Zehan Wang, Ruiqi Li, et~al.
\newblock Wavtokenizer: an efficient acoustic discrete codec tokenizer for audio language modeling.
\newblock \emph{arXiv preprint arXiv:2408.16532}, 2024.

\bibitem[Jiang et~al.(2023)Jiang, Peng, Xue, Zhang, and Lu]{jiang2023latent}
Xue Jiang, Xiulian Peng, Huaying Xue, Yuan Zhang, and Yan Lu.
\newblock Latent-domain predictive neural speech coding.
\newblock \emph{IEEE/ACM Transactions on Audio, Speech, and Language Processing}, 31:\penalty0 2111--2123, 2023.

\bibitem[Jiang et~al.(2024)Jiang, Liu, Ren, He, Ye, Ji, Yang, Zhang, Wei, Wang, et~al.]{jiang2024mega}
Ziyue Jiang, Jinglin Liu, Yi~Ren, Jinzheng He, Zhenhui Ye, Shengpeng Ji, Qian Yang, Chen Zhang, Pengfei Wei, Chunfeng Wang, et~al.
\newblock Mega-tts 2: Boosting prompting mechanisms for zero-shot speech synthesis.
\newblock In \emph{The Twelfth International Conference on Learning Representations}, 2024.

\bibitem[Jiang et~al.(2025)Jiang, Ren, Li, Ji, Zhang, Ye, Zhang, Jionghao, Yang, Zuo, et~al.]{jiang2025megatts}
Ziyue Jiang, Yi~Ren, Ruiqi Li, Shengpeng Ji, Boyang Zhang, Zhenhui Ye, Chen Zhang, Bai Jionghao, Xiaoda Yang, Jialong Zuo, et~al.
\newblock Megatts 3: Sparse alignment enhanced latent diffusion transformer for zero-shot speech synthesis.
\newblock \emph{arXiv preprint arXiv:2502.18924}, 2025.

\bibitem[Ju et~al.(2025)Ju, Yang, Yu, Shen, Leng, Wang, Tan, Zhou, Qin, and Li]{ju2025mooncast}
Zeqian Ju, Dongchao Yang, Jianwei Yu, Kai Shen, Yichong Leng, Zhengtao Wang, Xu~Tan, Xinyu Zhou, Tao Qin, and Xiangyang Li.
\newblock Mooncast: High-quality zero-shot podcast generation.
\newblock \emph{arXiv preprint arXiv:2503.14345}, 2025.

\bibitem[Kang et~al.(2024)Kang, Yang, Yao, Kuang, Yang, Guo, Lin, and Povey]{kang2024libriheavy}
Wei Kang, Xiaoyu Yang, Zengwei Yao, Fangjun Kuang, Yifan Yang, Liyong Guo, Long Lin, and Daniel Povey.
\newblock Libriheavy: A 50,000 hours asr corpus with punctuation casing and context.
\newblock In \emph{ICASSP 2024-2024 IEEE International Conference on Acoustics, Speech and Signal Processing (ICASSP)}, pages 10991--10995. IEEE, 2024.

\bibitem[Kim et~al.(2019)Kim, Kim, Lee, and Kim]{kim2019audiocaps}
Chris~Dongjoo Kim, Byeongchang Kim, Hyunmin Lee, and Gunhee Kim.
\newblock {AudioCaps}: Generating captions for audios in the wild.
\newblock In \emph{Proceedings of the 2019 Conference of the North American Chapter of the Association for Computational Linguistics: Human Language Technologies}, pages 119--132. Association for Computational Linguistics, 2019.
\newblock \doi{10.18653/v1/N19-1011}.

\bibitem[Kiritchenko and Mohammad(2017)]{kiritchenko2017bestworst}
Svetlana Kiritchenko and Saif Mohammad.
\newblock Best-worst scaling more reliable than rating scales: A case study on sentiment intensity annotation.
\newblock In \emph{Proceedings of the 55th Annual Meeting of the Association for Computational Linguistics (Volume 2: Short Papers)}, pages 465--470, Vancouver, Canada, 2017. Association for Computational Linguistics.

\bibitem[Kong et~al.(2020)Kong, Kim, and Bae]{kong2020hifi}
Jungil Kong, Jaehyeon Kim, and Jaekyoung Bae.
\newblock Hifi-gan: Generative adversarial networks for efficient and high fidelity speech synthesis.
\newblock \emph{Advances in neural information processing systems}, 33:\penalty0 17022--17033, 2020.

\bibitem[Kubichek(1993)]{kubichek1993mcd}
Robert~F. Kubichek.
\newblock Mel-cepstral distance measure for objective speech quality assessment.
\newblock In \emph{Proceedings of the IEEE Pacific Rim Conference on Communications, Computers and Signal Processing}, volume~1, pages 125--128, 1993.
\newblock \doi{10.1109/PACRIM.1993.407206}.

\bibitem[Kumar et~al.(2023{\natexlab{a}})Kumar, Tan, Ni, Manocha, Zhang, Henderson, and Xu]{kumar2023torchaudiosquim}
Anurag Kumar, Ke~Tan, Zhaoheng Ni, Pranay Manocha, Xiaohui Zhang, Ethan Henderson, and Buye Xu.
\newblock Torchaudio-squim: Reference-less speech quality and intelligibility measures in torchaudio.
\newblock In \emph{ICASSP 2023 - 2023 IEEE International Conference on Acoustics, Speech and Signal Processing (ICASSP)}, pages 1--5, 2023{\natexlab{a}}.

\bibitem[Kumar et~al.(2019)Kumar, Peng, and Levine]{kumar2019rewardconditioned}
Aviral Kumar, Xue~Bin Peng, and Sergey Levine.
\newblock Reward-conditioned policies.
\newblock \emph{arXiv preprint arXiv:1912.13465}, 2019.
\newblock URL \url{https://arxiv.org/abs/1912.13465}.

\bibitem[Kumar et~al.(2023{\natexlab{b}})Kumar, Seetharaman, Luebs, Kumar, and Kumar]{kumar2023dac}
Rithesh Kumar, Prem Seetharaman, Alejandro Luebs, Ishaan Kumar, and Kundan Kumar.
\newblock High-fidelity audio compression with improved rvqgan.
\newblock \emph{arXiv preprint arXiv:2306.06546}, 2023{\natexlab{b}}.

\bibitem[Lacombe et~al.(2024)Lacombe, Srivastav, and Gandhi]{lacombe2024parlertts}
Yoach Lacombe, Vaibhav Srivastav, and Sanchit Gandhi.
\newblock {Parler-TTS}.
\newblock \emph{GitHub repository}, 2024.
\newblock URL \url{https://github.com/huggingface/parler-tts}.

\bibitem[Lei et~al.(2026)Lei, Zhang, Pan, Pu, Guo, Li, and Zhao]{lei2026swansphere}
Ke~Lei, Yu~Zhang, Changhao Pan, Xueyi Pu, Wenxiang Guo, Ruiqi Li, and Zhou Zhao.
\newblock Towards streaming synchronized spatial audio generation via autoregressive diffusion transformer.
\newblock In \emph{Proceedings of the 43rd International Conference on Machine Learning}, 2026.

\bibitem[Li et~al.(2024)Li, Zhang, Wang, Hong, Huang, and Zhao]{li2024robust}
Ruiqi Li, Yu~Zhang, Yongqi Wang, Zhiqing Hong, Rongjie Huang, and Zhou Zhao.
\newblock Robust singing voice transcription serves synthesis.
\newblock \emph{arXiv preprint arXiv:2405.09940}, 2024.

\bibitem[Li et~al.(2026)Li, Zhang, Pan, Lei, Yin, and Yang]{li2026swanvoice}
Ruiqi Li, Yu~Zhang, Changhao Pan, Ke~Lei, Xiang Yin, and Cheng Yang.
\newblock {SwanVoice}: Expressive long-form zero-shot speech synthesis for both monologue and dialogue.
\newblock \emph{arXiv preprint arXiv:2605.30993}, 2026.

\bibitem[Lipman et~al.(2022)Lipman, Chen, Ben-Hamu, Nickel, and Le]{lipman2022flow}
Yaron Lipman, Ricky~TQ Chen, Heli Ben-Hamu, Maximilian Nickel, and Matthew Le.
\newblock Flow matching for generative modeling.
\newblock In \emph{The Eleventh International Conference on Learning Representations}, 2022.

\bibitem[Liu et~al.(2025)Liu, Li, Zhang, Teng, Jiang, Chen, Shi, Li, Wang, Chen, Meng, Zhao, Xu, He, Hu, and Zhang]{liu2025unimoeaudio}
Zhenyu Liu, Yunxin Li, Xuanyu Zhang, Qixun Teng, Shenyuan Jiang, Xinyu Chen, Haoyuan Shi, Jinchao Li, Qi~Wang, Haolan Chen, Fanbo Meng, Mingjun Zhao, Yu~Xu, Yancheng He, Baotian Hu, and Min Zhang.
\newblock {UniMoE-Audio}: Unified speech and music generation with dynamic-capacity {MoE}.
\newblock \emph{arXiv preprint arXiv:2510.13344}, 2025.

\bibitem[Loshchilov and Hutter(2019)]{loshchilov2019decoupled}
Ilya Loshchilov and Frank Hutter.
\newblock Decoupled weight decay regularization.
\newblock In \emph{International Conference on Learning Representations}, 2019.

\bibitem[Lyth and King(2024)]{lyth2024natural}
Dan Lyth and Simon King.
\newblock Natural language guidance of high-fidelity text-to-speech with synthetic annotations.
\newblock \emph{arXiv preprint arXiv:2402.01912}, 2024.

\bibitem[{MOSI}(2026)]{mosi_moss_ttsd_2026}
{MOSI}.
\newblock {MOSS-TTSD}.
\newblock \emph{Project website}, February 2026.
\newblock URL \url{https://mosi.cn/models/moss-ttsd}.

\bibitem[Oord et~al.(2018)Oord, Li, and Vinyals]{oord2018representation}
Aaron van~den Oord, Yazhe Li, and Oriol Vinyals.
\newblock Representation learning with contrastive predictive coding.
\newblock \emph{arXiv preprint arXiv:1807.03748}, 2018.

\bibitem[Pan et~al.(2025)Pan, Guo, Zhang, Zhu, Chen, Wang, and Zhao]{pan2025spatialeval}
Changhao Pan, Wenxiang Guo, Yu~Zhang, Zhiyuan Zhu, Zhetao Chen, Han Wang, and Zhou Zhao.
\newblock A multimodal evaluation framework for spatial audio playback systems: From localization to listener preference.
\newblock In \emph{Proceedings of the 33rd ACM International Conference on Multimedia}, pages 7006--7015. ACM, 2025.

\bibitem[Pan et~al.(2026{\natexlab{a}})Pan, Fan, Zhuo, Chen, Guo, Zhang, Li, Zhu, Yang, Ji, Wen, Xu, Lei, Yang, Lu, and Zhao]{pan2026audioediting}
Changhao Pan, Yifei Fan, Fan Zhuo, Yifu Chen, Wenxiang Guo, Yu~Zhang, Ruiqi Li, Zhiyuan Zhu, Rui Yang, Shengpeng Ji, Chenyuhao Wen, Jiayang Xu, Ke~Lei, Xiaoda Yang, Jingyu Lu, and Zhou Zhao.
\newblock Audio editing in the era of foundation models: A survey.
\newblock \emph{arXiv preprint arXiv:2606.23139}, 2026{\natexlab{a}}.
\newblock URL \url{https://arxiv.org/abs/2606.23139}.

\bibitem[Pan et~al.(2026{\natexlab{b}})Pan, Yang, Wang, Zhou, He, Guo, Jiang, Li, Zhang, Wen, Lei, Yin, Lu, Zhu, and Zhao]{pan2026swanbench}
Changhao Pan, Rui Yang, Han Wang, Zhuan Zhou, Xuming He, Wenxiang Guo, Ziyue Jiang, Ruiqi Li, Yu~Zhang, Chenyuhao Wen, Ke~Lei, Xiang Yin, Jingyu Lu, Zhiyuan Zhu, and Zhou Zhao.
\newblock Comprehensive benchmarking of long-form speech generation in diverse scenarios.
\newblock \emph{arXiv preprint arXiv:2605.28618}, 2026{\natexlab{b}}.

\bibitem[Panayotov et~al.(2015)Panayotov, Chen, Povey, and Khudanpur]{panayotov2015librispeech}
Vassil Panayotov, Guoguo Chen, Daniel Povey, and Sanjeev Khudanpur.
\newblock Librispeech: an asr corpus based on public domain audio books.
\newblock In \emph{2015 IEEE international conference on acoustics, speech and signal processing (ICASSP)}, pages 5206--5210. IEEE, 2015.

\bibitem[Parker et~al.(2026)Parker, Evans, Carr, Zukowski, Taylor, Rice, and Pons]{parker2026same}
Julian~D Parker, Zach Evans, CJ~Carr, Zachary Zukowski, Josiah Taylor, Matthew Rice, and Jordi Pons.
\newblock Same: A semantically-aligned music autoencoder.
\newblock \emph{arXiv preprint arXiv:2605.18613}, 2026.

\bibitem[Peebles and Xie(2023)]{peebles2023scalable}
William Peebles and Saining Xie.
\newblock Scalable diffusion models with transformers.
\newblock In \emph{Proceedings of the IEEE/CVF International Conference on Computer Vision}, pages 4195--4205, 2023.

\bibitem[Peng et~al.(2025)Peng, Yu, Wang, Chang, Sun, Dong, Zhu, Xu, Bao, Wang, Huang, Xia, and Wei]{peng2025vibevoice}
Zhiliang Peng, Jianwei Yu, Wenhui Wang, Yaoyao Chang, Yutao Sun, Li~Dong, Yi~Zhu, Weijiang Xu, Hangbo Bao, Zehua Wang, Shaohan Huang, Yan Xia, and Furu Wei.
\newblock Vibevoice technical report.
\newblock \emph{arXiv preprint arXiv:2508.19205}, 2025.

\bibitem[{Qwen Team}(2026)]{qwen2026qwen3tts}
{Qwen Team}.
\newblock {Qwen3-TTS} technical report.
\newblock \emph{arXiv preprint arXiv:2601.15621}, 2026.

\bibitem[Rastorgueva et~al.(2023)Rastorgueva, Lavrukhin, and Ginsburg]{rastorgueva2023nemo}
Elena Rastorgueva, Vitaly Lavrukhin, and Boris Ginsburg.
\newblock Nemo forced aligner and its application to word alignment for subtitle generation.
\newblock In \emph{Interspeech}, pages 5257--5258, 2023.

\bibitem[Reddy et~al.(2021)Reddy, Gopal, and Cutler]{reddy2021dnsmos}
Chandan K.~A. Reddy, Vishak Gopal, and Ross Cutler.
\newblock Dnsmos: A non-intrusive perceptual objective speech quality metric to evaluate noise suppressors.
\newblock In \emph{ICASSP 2021 - 2021 IEEE International Conference on Acoustics, Speech and Signal Processing (ICASSP)}, pages 6493--6497, 2021.

\bibitem[Ren et~al.(2026)Ren, Yi, Tao, Sun, Wen, Gu, Xu, and Bai]{ren2026ovinstructtts}
Yong Ren, Jiangyan Yi, Jianhua Tao, Haiyang Sun, Zhengqi Wen, Hao Gu, Le~Xu, and Ye~Bai.
\newblock {OV-InstructTTS}: Towards open-vocabulary instruct text-to-speech.
\newblock \emph{arXiv preprint arXiv:2601.01459}, 2026.

\bibitem[Rix et~al.(2001)Rix, Beerends, Hollier, and Hekstra]{rix2001pesq}
A.~W. Rix, J.~G. Beerends, M.~P. Hollier, and A.~P. Hekstra.
\newblock Perceptual evaluation of speech quality (pesq)-a new method for speech quality assessment of telephone networks and codecs.
\newblock In \emph{2001 IEEE International Conference on Acoustics, Speech, and Signal Processing (ICASSP)}, volume~2, pages 749--752, 2001.

\bibitem[Shao et~al.(2024)Shao, Wang, Zhu, Xu, Song, Bi, Zhang, Zhang, Li, Wu, and Guo]{shao2024deepseekmath}
Zhihong Shao, Peiyi Wang, Qihao Zhu, Runxin Xu, Junxiao Song, Xiao Bi, Haowei Zhang, Mingchuan Zhang, Y.~K. Li, Y.~Wu, and Daya Guo.
\newblock {DeepSeekMath}: Pushing the limits of mathematical reasoning in open language models.
\newblock \emph{arXiv preprint arXiv:2402.03300}, 2024.

\bibitem[Shen et~al.(2023)Shen, Ju, Tan, Liu, Leng, He, Qin, Zhao, and Bian]{shen2023naturalspeech}
Kai Shen, Zeqian Ju, Xu~Tan, Yanqing Liu, Yichong Leng, Lei He, Tao Qin, Sheng Zhao, and Jiang Bian.
\newblock Naturalspeech 2: Latent diffusion models are natural and zero-shot speech and singing synthesizers.
\newblock \emph{arXiv preprint arXiv:2304.09116}, 2023.

\bibitem[Shi et~al.(2020)Shi, Bu, Xu, Zhang, and Li]{shi2020aishell}
Yao Shi, Hui Bu, Xin Xu, Shaoji Zhang, and Ming Li.
\newblock Aishell-3: A multi-speaker mandarin tts corpus and the baselines.
\newblock \emph{arXiv preprint arXiv:2010.11567}, 2020.

\bibitem[Skorokhodov et~al.(2025)Skorokhodov, Girish, Hu, Menapace, Li, Abdal, Tulyakov, and Siarohin]{skorokhodov2025diffusability}
Ivan Skorokhodov, Sharath Girish, Benran Hu, Willi Menapace, Yanyu Li, Rameen Abdal, Sergey Tulyakov, and Aliaksandr Siarohin.
\newblock Improving the diffusability of autoencoders.
\newblock \emph{arXiv preprint arXiv:2502.14831}, 2025.

\bibitem[St{\"o}ter et~al.(2018)St{\"o}ter, Liutkus, and Ito]{stoter2018sisec}
Fabian-Robert St{\"o}ter, Antoine Liutkus, and Nobutaka Ito.
\newblock The 2018 signal separation evaluation campaign.
\newblock In \emph{Latent Variable Analysis and Signal Separation}, pages 293--305. Springer, 2018.

\bibitem[Taal et~al.(2011)Taal, Hendriks, Heusdens, and Jensen]{taal2011algorithm}
Cees~H Taal, Richard~C Hendriks, Richard Heusdens, and Jesper Jensen.
\newblock An algorithm for intelligibility prediction of time--frequency weighted noisy speech.
\newblock \emph{IEEE Transactions on audio, speech, and language processing}, 19\penalty0 (7):\penalty0 2125--2136, 2011.

\bibitem[{Team Seedance} et~al.(2026){Team Seedance}, Chen, Chen, Chen, et~al.]{teamseedance2026seedance2}
{Team Seedance}, De~Chen, Liyang Chen, Xin Chen, et~al.
\newblock {Seedance 2.0}: Advancing video generation for world complexity.
\newblock \emph{arXiv preprint arXiv:2604.14148}, 2026.

\bibitem[Wang et~al.(2026)Wang, Tian, Li, Lv, Zhao, and Li]{wang2026flowttsgrpo}
Haoxu Wang, Biao Tian, Weiqin Li, Xiang Lv, Han Zhao, and Xiangang Li.
\newblock Flowtts-grpo: Online reinforcement learning with multi-objective reward optimization for flow-matching based text-to-speech.
\newblock \emph{arXiv preprint arXiv:2606.23190}, 2026.

\bibitem[Wang et~al.(2023)Wang, Zheng, Chen, Cheng, and Chen]{wang2023campp}
Hui Wang, Siqi Zheng, Yafeng Chen, Luyao Cheng, and Qian Chen.
\newblock Cam++: A fast and efficient network for speaker verification using context-aware masking.
\newblock \emph{arXiv preprint arXiv:2303.00332}, 2023.

\bibitem[Wang et~al.(2024)Wang, Gao, Zhao, Sun, and Dai]{wang2024auxiliarylossfree}
Lean Wang, Huazuo Gao, Chenggang Zhao, Xu~Sun, and Damai Dai.
\newblock Auxiliary-loss-free load balancing strategy for mixture-of-experts.
\newblock \emph{arXiv preprint arXiv:2408.15664}, 2024.

\bibitem[Wang et~al.(2025)Wang, Jiang, Ma, Zhang, Liu, Li, Liang, Zheng, Wang, Feng, et~al.]{wang2025spark}
Xinsheng Wang, Mingqi Jiang, Ziyang Ma, Ziyu Zhang, Songxiang Liu, Linqin Li, Zheng Liang, Qixi Zheng, Rui Wang, Xiaoqin Feng, et~al.
\newblock Spark-tts: An efficient llm-based text-to-speech model with single-stream decoupled speech tokens.
\newblock \emph{arXiv preprint arXiv:2503.01710}, 2025.

\bibitem[Xie et~al.(2025{\natexlab{a}})Xie, Lin, Cao, Guo, Tian, Wu, Wen, Shang, Liu, Jiang, et~al.]{xie2025soulx}
Hanke Xie, Haopeng Lin, Wenxiao Cao, Dake Guo, Wenjie Tian, Jun Wu, Hanlin Wen, Ruixuan Shang, Hongmei Liu, Zhiqi Jiang, et~al.
\newblock Soulx-podcast: Towards realistic long-form podcasts with dialectal and paralinguistic diversity.
\newblock \emph{arXiv preprint arXiv:2510.23541}, 2025{\natexlab{a}}.

\bibitem[Xie et~al.(2025{\natexlab{b}})Xie, Shen, Li, Xie, Tang, and Hu]{xie2025fireredtts}
Kun Xie, Feiyu Shen, Junjie Li, Fenglong Xie, Xu~Tang, and Yao Hu.
\newblock Fireredtts-2: Towards long conversational speech generation for podcast and chatbot.
\newblock \emph{arXiv preprint arXiv:2509.02020}, 2025{\natexlab{b}}.

\bibitem[Xu et~al.(2024)Xu, Chen, Yu, Huang, Wu, Zhang, Li, Luo, and Gu]{xu2024secap}
Yaoxun Xu, Hangting Chen, Jianwei Yu, Qiaochu Huang, Zhiyong Wu, Shi-Xiong Zhang, Guangzhi Li, Yi~Luo, and Rongzhi Gu.
\newblock Secap: Speech emotion captioning with large language model.
\newblock In \emph{Proceedings of the AAAI Conference on Artificial Intelligence}, volume~38, pages 19323--19331, 2024.

\bibitem[Yamagishi et~al.(2019)Yamagishi, Veaux, and MacDonald]{yamagishi2019vctk}
Junichi Yamagishi, Christophe Veaux, and Kirsten MacDonald.
\newblock {CSTR VCTK Corpus}: English multi-speaker corpus for {CSTR} voice cloning toolkit (version 0.92), 2019.

\bibitem[Yang et~al.(2025)Yang, Li, Yang, et~al.]{yang2025qwen3}
An~Yang, Anfeng Li, Baosong Yang, et~al.
\newblock {Qwen3} technical report.
\newblock \emph{arXiv preprint arXiv:2505.09388}, 2025.

\bibitem[Yang et~al.(2024)Yang, Quan, Wang, Wang, Yang, Fang, Shao, Bu, Xu, and Li]{yang2024realman}
Bing Yang, Changsheng Quan, Yabo Wang, Pengyu Wang, Yujie Yang, Ying Fang, Nian Shao, Hui Bu, Xin Xu, and Xiaofei Li.
\newblock Realman: A real-recorded and annotated microphone array dataset for dynamic speech enhancement and localization.
\newblock \emph{Advances in Neural Information Processing Systems}, 37:\penalty0 105997--106019, 2024.

\bibitem[Yang et~al.(2023)Yang, Tian, Tan, Huang, Liu, Chang, Shi, Zhao, Bian, Wu, et~al.]{yang2023uniaudio}
Dongchao Yang, Jinchuan Tian, Xu~Tan, Rongjie Huang, Songxiang Liu, Xuankai Chang, Jiatong Shi, Sheng Zhao, Jiang Bian, Xixin Wu, et~al.
\newblock Uniaudio: An audio foundation model toward universal audio generation.
\newblock \emph{arXiv preprint arXiv:2310.00704}, 2023.

\bibitem[Zen et~al.(2019)Zen, Dang, Clark, Zhang, Weiss, Jia, Chen, and Wu]{zen2019libritts}
Heiga Zen, Viet Dang, Rob Clark, Yu~Zhang, Ron~J. Weiss, Ye~Jia, Zhifeng Chen, and Yonghui Wu.
\newblock Libritts: A corpus derived from librispeech for text-to-speech.
\newblock In \emph{Proc. Interspeech}, pages 1526--1530, 2019.

\bibitem[Zezario et~al.(2020)Zezario, Fu, Fuh, Tsao, and Wang]{zezario2020stoi}
Ryandhimas~E. Zezario, Szu-Wei Fu, Chiou-Shann Fuh, Yu~Tsao, and Hsin-Min Wang.
\newblock Stoi-net: {A} deep learning based non-intrusive speech intelligibility assessment model.
\newblock In \emph{Asia-Pacific Signal and Information Processing Association Annual Summit and Conference, {APSIPA} 2020, Auckland, New Zealand, December 7--10, 2020}, pages 482--486. {IEEE}, 2020.

\bibitem[Zhang and Sennrich(2019)]{zhang2019root}
Biao Zhang and Rico Sennrich.
\newblock Root mean square layer normalization.
\newblock \emph{Advances in Neural Information Processing Systems}, 32, 2019.

\bibitem[Zhang et~al.(2024{\natexlab{a}})Zhang, Qian, Zhou, Liu, Wang, Wang, Yousefi, Qian, Li, He, et~al.]{zhang2024covomix}
Leying Zhang, Yao Qian, Long Zhou, Shujie Liu, Dongmei Wang, Xiaofei Wang, Midia Yousefi, Yanmin Qian, Jinyu Li, Lei He, et~al.
\newblock Covomix: Advancing zero-shot speech generation for human-like multi-talker conversations.
\newblock \emph{Advances in Neural Information Processing Systems}, 37:\penalty0 100291--100317, 2024{\natexlab{a}}.

\bibitem[Zhang(2019)]{zhang2019making}
Richard Zhang.
\newblock Making convolutional networks shift-invariant again.
\newblock In \emph{International conference on machine learning}, pages 7324--7334. PMLR, 2019.

\bibitem[Zhang et~al.(2025{\natexlab{a}})Zhang, Liu, Li, Li, Wang, Zhu, Wang, et~al.]{zhang2025mimoaudio}
Xin Zhang, Tianrui Liu, Han Li, Zhi Li, Wenbo Wang, Yunhua Zhu, Zheng Wang, et~al.
\newblock {MiMo-Audio}: Audio language models are few-shot learners.
\newblock \emph{arXiv preprint arXiv:2512.23808}, 2025{\natexlab{a}}.

\bibitem[Zhang et~al.(2026)Zhang, Wang, Liao, Li, Wang, Wang, Jia, Chen, Li, Chen, and Wu]{zhang2026speechjudge}
Xueyao Zhang, Chaoren Wang, Huan Liao, Ziniu Li, Yuancheng Wang, Li~Wang, Dongya Jia, Yuanzhe Chen, Xiulin Li, Zhuo Chen, and Zhizheng Wu.
\newblock {SpeechJudge}: Towards human-level judgment for speech naturalness.
\newblock In \emph{International Conference on Learning Representations}, 2026.

\bibitem[Zhang et~al.(2024{\natexlab{b}})Zhang, Huang, Li, He, Xia, Chen, Duan, Huai, and Zhao]{zhang2024stylesinger}
Yu~Zhang, Rongjie Huang, Ruiqi Li, JinZheng He, Yan Xia, Feiyang Chen, Xinyu Duan, Baoxing Huai, and Zhou Zhao.
\newblock Stylesinger: Style transfer for out-of-domain singing voice synthesis.
\newblock In \emph{Proceedings of the AAAI Conference on Artificial Intelligence}, volume~38, pages 19597--19605, 2024{\natexlab{b}}.

\bibitem[Zhang et~al.(2024{\natexlab{c}})Zhang, Jiang, Li, Pan, He, Huang, Wang, and Zhao]{zhang2024tcsinger}
Yu~Zhang, Ziyue Jiang, Ruiqi Li, Changhao Pan, Jinzheng He, Rongjie Huang, Chuxin Wang, and Zhou Zhao.
\newblock Tcsinger: Zero-shot singing voice synthesis with style transfer and multi-level style control.
\newblock In \emph{Proceedings of the 2024 Conference on Empirical Methods in Natural Language Processing}, pages 1960--1975, 2024{\natexlab{c}}.

\bibitem[Zhang et~al.(2024{\natexlab{d}})Zhang, Pan, Guo, Li, Zhu, Wang, Xu, Lu, Hong, Wang, et~al.]{zhang2024gtsinger}
Yu~Zhang, Changhao Pan, Wenxiang Guo, Ruiqi Li, Zhiyuan Zhu, Jialei Wang, Wenhao Xu, Jingyu Lu, Zhiqing Hong, Chuxin Wang, et~al.
\newblock Gtsinger: A global multi-technique singing corpus with realistic music scores for all singing tasks.
\newblock \emph{Advances in Neural Information Processing Systems (NeurIPS)}, 2024{\natexlab{d}}.

\bibitem[Zhang et~al.(2025{\natexlab{b}})Zhang, Guo, Pan, Yao, Zhu, Jiang, Wang, Jin, and Zhao]{zhang2025tcsinger}
Yu~Zhang, Wenxiang Guo, Changhao Pan, Dongyu Yao, Zhiyuan Zhu, Ziyue Jiang, Yuhan Wang, Tao Jin, and Zhou Zhao.
\newblock {TCSinger} 2: Customizable multilingual zero-shot singing voice synthesis.
\newblock In \emph{Proceedings of the 63rd Annual Meeting of the Association for Computational Linguistics}, pages 13280--13294, Vienna, Austria, 2025{\natexlab{b}}. Association for Computational Linguistics.

\bibitem[Zhang et~al.(2025{\natexlab{c}})Zhang, Guo, Pan, Zhu, Jin, and Zhao]{zhang2025isdrama}
Yu~Zhang, Wenxiang Guo, Changhao Pan, Zhiyuan Zhu, Tao Jin, and Zhou Zhao.
\newblock Isdrama: Immersive spatial drama generation through multimodal prompting.
\newblock \emph{arXiv preprint arXiv:2504.20630}, 2025{\natexlab{c}}.

\bibitem[Zhang et~al.(2025{\natexlab{d}})Zhang, Guo, Pan, Zhu, Li, Lu, Huang, Zhang, Hong, Jiang, et~al.]{zhang2025versatile}
Yu~Zhang, Wenxiang Guo, Changhao Pan, Zhiyuan Zhu, Ruiqi Li, Jingyu Lu, Rongjie Huang, Ruiyuan Zhang, Zhiqing Hong, Ziyue Jiang, et~al.
\newblock Versatile framework for song generation with prompt-based control.
\newblock \emph{arXiv preprint arXiv:2504.19062}, 2025{\natexlab{d}}.

\bibitem[Zhang et~al.(2025{\natexlab{e}})Zhang, Tian, and Duan]{zhang2025conan}
Yu~Zhang, Baotong Tian, and Zhiyao Duan.
\newblock {Conan}: A chunkwise online network for zero-shot adaptive voice conversion.
\newblock In \emph{2025 IEEE Automatic Speech Recognition and Understanding Workshop}, 2025{\natexlab{e}}.

\bibitem[Zhou et~al.(2025)Zhou, Zhou, He, Zhou, Wang, Deng, and Shu]{zhou2025indextts2}
Siyi Zhou, Yiquan Zhou, Yi~He, Xun Zhou, Jinchao Wang, Wei Deng, and Jingchen Shu.
\newblock Indextts2: A breakthrough in emotionally expressive and duration-controlled auto-regressive zero-shot text-to-speech.
\newblock \emph{arXiv preprint arXiv:2506.21619}, 2025.

\bibitem[Zhou et~al.(2024)Zhou, Qin, Jin, Zhou, Lei, Zhou, Wu, and Jia]{zhou2024voxinstruct}
Yixuan Zhou, Xiaoyu Qin, Zeyu Jin, Shuoyi Zhou, Shun Lei, Songtao Zhou, Zhiyong Wu, and Jia Jia.
\newblock {VoxInstruct}: Expressive human instruction-to-speech generation with unified multilingual codec language modelling.
\newblock In \emph{Proceedings of the 32nd ACM International Conference on Multimedia}, pages 554--563, 2024.

\bibitem[Zhou et~al.(2026)Zhou, Zeng, Liu, Li, Yu, Gui, Wu, Wang, Shen, Ye, Zhang, Zhou, Bai, Sun, Deng, Shi, Wu, and Liu]{zhou2026voxcpm2}
Yixuan Zhou, Guoyang Zeng, Xin Liu, Xiang Li, Renjie Yu, Jiancheng Gui, Jiaheng Wu, Ziyang Wang, Xudong Shen, Runchuan Ye, Zhisheng Zhang, Jiuyang Zhou, Bingsong Bai, Weiyue Sun, Mengyuan Deng, Qundong Shi, Zhiyong Wu, and Zhiyuan Liu.
\newblock {VoxCPM2} technical report.
\newblock \emph{arXiv preprint arXiv:2606.06928}, 2026.

\bibitem[Zhu et~al.(2025{\natexlab{a}})Zhu, Kang, Guo, Yao, Kuang, Zhuang, Li, Han, Zhang, Zhang, et~al.]{zhu2025zipvoice}
Han Zhu, Wei Kang, Liyong Guo, Zengwei Yao, Fangjun Kuang, Weiji Zhuang, Zhaoqing Li, Zhifeng Han, Dong Zhang, Xin Zhang, et~al.
\newblock Zipvoice-dialog: Non-autoregressive spoken dialogue generation with flow matching.
\newblock \emph{arXiv preprint arXiv:2507.09318}, 2025{\natexlab{a}}.

\bibitem[Zhu et~al.(2025{\natexlab{b}})Zhu, Zhang, Guo, Pan, and Zhao]{zhu2025asaudio}
Zhiyuan Zhu, Yu~Zhang, Wenxiang Guo, Changhao Pan, and Zhou Zhao.
\newblock {ASAudio}: A survey of advanced spatial audio research.
\newblock In \emph{Proceedings of the 14th International Joint Conference on Natural Language Processing and the 4th Conference of the Asia-Pacific Chapter of the Association for Computational Linguistics}, pages 417--442, Mumbai, India, 2025{\natexlab{b}}. The Asian Federation of Natural Language Processing and The Association for Computational Linguistics.

\bibitem[Zoph et~al.(2022)Zoph, Bello, Kumar, Du, Huang, Dean, Shazeer, and Fedus]{zoph2022stmoe}
Barret Zoph, Irwan Bello, Sameer Kumar, Nan Du, Yanping Huang, Jeff Dean, Noam Shazeer, and William Fedus.
\newblock {ST-MoE}: Designing stable and transferable sparse expert models.
\newblock \emph{arXiv preprint arXiv:2202.08906}, 2022.

\end{thebibliography}

\clearpage
\beginappendix
\section{Caption Style Matrices}
\label{apx:caption_style_matrix}

\noindent\textbf{Scope.}
The matrices below condense the annotation guides for three media families introduced in Section~\ref{sec:data}: animation; short drama and film/TV drama; and advertisement and digital-human content. The source guides are hierarchical: animation moves from broad audience and topic trends to role-level vocal archetypes, and drama refines role identities with dialogue and delivery evidence. Advertisement and digital-human annotation combine persona and voice-style candidates with expressiveness and industry-specific speaking strategies. The matrices provide soft priors for clips that follow strong media conventions and are not used for ordinary data.

\noindent\textbf{Annotation procedure.}
Annotators first select the applicable matrix using the scene trigger. Only speakers who actually speak are listed, ordered by their first utterance. Each \texttt{Speakers} entry starts with perceived gender and age range and adds three to five stable, discriminative characteristics: a role or persona when supported by speech or delivery, stable timbre, and habitual delivery. The matrix supplies candidate descriptions for these characteristics. Utterance-level changes in emotion, pace, loudness, pausing, emphasis, hesitation, and local audio effects are recorded chronologically in \texttt{Content}. Every retained descriptor must be supported by audible evidence at the relevant temporal location in the recording.

\begin{table}[!htbp]
  \centering
  \caption{Condensed style matrix for animation-style captions.}
  \label{tab:style_matrix_animation}
  \begin{tabular}{p{0.23\linewidth}p{0.67\linewidth}}
    \toprule
    Aspect & Condensed rule \\
    \midrule
    Typical triggers & Animation, cartoon, anime, dubbing, role-playing voices, and other clips whose delivery follows a character-dubbing convention. \\
    Stable speaker profile & Describe perceived gender, approximate age, an audible vocal archetype when useful, stable timbre, and habitual delivery, in that order. Archetypes such as an energetic lead, a restrained mature speaker, or a comic supporting voice require clear evidence in the vocal performance. \\
    Local delivery & Record exaggerated reactions, abrupt emotional shifts, punch-line timing, shouts, laughter, hesitation, and changes in pace, loudness, or arousal in the chronological \texttt{Content} field. \\
    Acoustic evidence & Ground descriptions in cues such as habitual pitch range, brightness, breathiness, energy, attack strength, pausing, and the degree of restraint or exaggeration. \\
    Representative distinctions & Action-oriented clips favor larger loudness dynamics, faster pace, and stronger bursts; romance favors finer emotional control, breathiness, and pauses; suspense favors restrained, clear delivery; historical or courtly settings favor formal diction and measured expression. \\
    \bottomrule
  \end{tabular}
\end{table}

\begin{table}[!htbp]
  \centering
  \caption{Condensed style matrix for short-drama and film/TV-drama-style captions.}
  \label{tab:style_matrix_drama}
  \begin{tabular}{p{0.23\linewidth}p{0.67\linewidth}}
    \toprule
    Aspect & Condensed rule \\
    \midrule
    Typical triggers & Short drama, micro drama, vertical drama, scripted short video, web drama, film, TV drama, and other dialogue-heavy staged media. \\
    Stable speaker profile & Describe perceived gender, approximate age, a role or social identity supported by spoken dialogue or vocal delivery, stable timbre, and habitual delivery. \\
    Local delivery & Record interruption, conflict, emotional escalation, reversal, pleading, threat, command, hesitation, and relationship-driven changes in pace, loudness, or tone in the chronological \texttt{Content} field. \\
    Acoustic evidence & Use audible properties such as pacing, diction, theatrical coloring, controlled pauses, coldness, ingratiating delivery, and abrupt changes in intensity to describe delivery. Character identity requires supporting dialogue or role evidence. \\
    Representative distinctions & Examples include secretary-like delivery (fast, clear, formal), guard-like delivery (steady, terse, forceful), ingratiating or eunuch-like delivery (thin voice, raised endings, deferential wording). \\
    \bottomrule
  \end{tabular}
\end{table}

\begin{table}[!htbp]
  \centering
  \caption{Condensed style matrix for advertisement and digital-human-style captions.}
  \label{tab:style_matrix_ad_digitalhuman}
  \begin{tabular}{p{0.23\linewidth}p{0.67\linewidth}}
    \toprule
    Aspect & Condensed rule \\
    \midrule
    Typical triggers & Advertisement, commercial voice-over, digital-human content, livestream selling, product recommendation, product seeding, marketing speech, and scripted promotional narration. \\
    Stable speaker profile & Describe perceived gender, approximate age, a persona type supported by the speech function, voice-style class, timbre, and habitual product-pitch delivery. Persona labels such as host, product recommender, lecturer, or service worker require evidence from the spoken content or delivery. \\
    Local delivery & Record selling-point emphasis, urgency, price or discount emphasis, calls to action, question hooks, trust-building explanations, conversational softening, and changes in excitement in the chronological \texttt{Content} field. \\
    Acoustic evidence & Ground descriptions in audible properties such as friendliness, authority, energy, technical density, conversational warmth, cadence regularity, pause timing, emotional range, and script-like phrasing. Synthetic-voice judgments require direct audible artifacts. \\
    Representative distinctions & Representative groups include product recommenders and livestream hosts, health or education explainers, finance or business speakers, and service roles; matching styles range from conversational sharing and storytelling to energetic sales and structured explanation. \\
    \bottomrule
  \end{tabular}
\end{table}

\section{SwanVerifier}
\label{apx:swanverifier}

\subsection{Motivation}

The \texttt{Speakers} field provides SwanTale with stable, controllable speaker attributes. Perceived age and gender are difficult for automatic captioners in noisy media audio, child or elderly speech, role-playing voices, advertisements, animation, and game-style dubbing. Because these labels recur throughout the training data, a systematic error would create repeated supervision rather than an isolated captioning mistake.

SwanVerifier provides a waveform-grounded check for these coarse attributes. It tests whether the demographic labels are acoustically plausible and abstains when the waveform or prediction is ambiguous; caption generation is outside its scope. Detailed persona, role, and expressive style are still handled by caption annotation and human audit whenever automatic evidence is insufficient.

\subsection{Overview}

SwanVerifier is a compact audio tagger built on a pretrained WavLM encoder~\citep{chen2022wavlm}. The input waveform is resampled to 16~kHz and encoded as frame-level speech representations, which are pooled by attribute-specific heads to produce utterance-level predictions. For its primary consistency checks, SwanData-Caption uses the following two demographic heads:
\begin{itemize}
    \item \textbf{Age group}: \texttt{Child}, \texttt{Teenager}, \texttt{Youth-Adult}, \texttt{Middle-aged}, and \texttt{Elderly}.
    \item \textbf{Perceived gender}: the \texttt{male} and \texttt{female} labels in the caption inventory.
\end{itemize}

The underlying tagger also exposes emotion, pitch, pitch standard deviation, and speaking speed. We treat these outputs as auxiliary evidence, not as grounds for automatic demographic correction. Transient affect, emphasis, hesitation, and scene-specific performance stay in \texttt{Content} and are checked during the final captioning and auditing pass for each retained sample.

\subsection{Problem Setup}

Let $c$ be a caption containing the following speaker inventory:
\[
\mathcal{D}(c)=\{d_1,\ldots,d_K\},
\]
where $d_k$ is the normalized description of speaker $k$ in the \texttt{Speakers} field. For verification, $x_k$ denotes a vocal segment attributed to that speaker. Automatic demographic checking is applied only when $x_k$ contains a single acoustically separable speaker; segments with unresolved overlap, cross-talk, or uncertain attribution bypass hard verification and are left to later auditing before any automatic repair.

When the corresponding tokens are present in $d_k$, let $y_{a,k}$ and $y_{g,k}$ denote its normalized age and perceived-gender labels. SwanVerifier estimates their class distributions as follows:
\[
p_a(\cdot\mid x_k), \qquad p_g(\cdot\mid x_k),
\]
and compares each sufficiently confident prediction with the corresponding caption label. Missing labels are not inferred or inserted. SwanVerifier is therefore a selective consistency check, not a complete speaker profiler.

\subsection{Backbone Encoding and Prediction Heads}

WavLM maps a vocal segment to frame-level hidden states,
\[
H_k=\mathrm{Enc}_{\mathrm{WavLM}}(x_k)\in\mathbb{R}^{T_k\times d},
\]
where $T_k$ is the number of valid acoustic frames. For each demographic attribute $r\in\{a,g\}$, an attention-pooling head forms the following utterance-level representation:
\[
\alpha^{(r)}_{k,t}=
\frac{\exp(q_r^\top h_{k,t})}
{\sum_{\tau=1}^{T_k}\exp(q_r^\top h_{k,\tau})},
\qquad
v_{k,r}=\sum_{t=1}^{T_k}\alpha^{(r)}_{k,t}h_{k,t}.
\]
The corresponding classifier produces the following output distribution:
\[
o_{k,r}=W_rv_{k,r}+b_r,
\qquad
p_r(\cdot\mid x_k)=\mathrm{softmax}(o_{k,r}).
\]
Separate heads allow the age and gender predictions to be calibrated and audited independently.

\subsection{Training Objective}

The demographic heads are trained with cross-entropy over normalized labels:
\[
\mathcal{L}_{\mathrm{tag}}
=\lambda_a\,w_{y_{a,k}}\,\mathrm{CE}\!\left(p_a(\cdot\mid x_k),y_{a,k}\right)
+\lambda_g\,\mathrm{CE}\!\left(p_g(\cdot\mid x_k),y_{g,k}\right),
\]
where $w_{y_{a,k}}$ compensates for age-class imbalance. A term is evaluated only when its label is available. The auxiliary emotion head is trained on samples with a non-unknown emotion label; the remaining auxiliary outputs are used only as secondary signals in this pipeline.

\subsection{Training and Scope}

We fine-tune an existing WavLM-based tagger on labeled speech whose speaker and utterance attributes can be mapped to the verifier taxonomy. Samples without an unambiguous acoustic subject are excluded from supervised demographic checking. SwanVerifier is an internal filtering component rather than a new tagging benchmark; the results below therefore characterize the verifier used in the data pipeline and should not be read as a comparison with dedicated speaker-attribute systems.

\subsection{Evaluation}

We evaluate the final model on a held-out labeled split. Table~\ref{tab:swanverifier_eval} reports utterance-level accuracy; emotion accuracy is computed only on samples with a non-unknown emotion label.

\begin{table}[!htbp]
  \centering
  \caption{Utterance-level accuracy (\%) of SwanVerifier on the held-out labeled split.}
  \label{tab:swanverifier_eval}
  \begin{tabular}{lccc}
    \toprule
    Split & Age & Gender & Emotion \\
    \midrule
    Held-out labeled split & 86.11 & 97.60 & 92.75 \\
    \bottomrule
  \end{tabular}
\end{table}

These results establish the operating accuracy of the filtering component on its held-out split. They do not show that every prediction is safe for automatic correction, so inference also requires confidence-based abstention and leaves ambiguous cases to manual audit.

\subsection{Inference Procedure}

During caption validation, SwanVerifier processes speaker-attributed segments from the vocal stream. A prediction is compared with a normalized demographic token only when it exceeds the confidence threshold for that attribute. A confident match leaves the caption unchanged; a confident mismatch flags the sample for repair, re-captioning, or removal. Low-confidence predictions, unresolved speaker overlap, and missing demographic tokens produce no automatic decision under these confidence-based validation rules.

Restricting automatic decisions in this way targets systematic age- and gender-label errors without treating inferred persona, role identity, or transient delivery as demographic evidence. Caption-level checks and human listening audits handle those richer attributes throughout the data pipeline.

\end{document}